\documentclass[conference,letterpaper]{IEEEtran}

\usepackage{cite}
\usepackage{amsmath,amssymb,amsfonts}
\usepackage{algorithmic}
\usepackage{graphicx}
\usepackage{textcomp}
\usepackage{xcolor}
\usepackage{xspace}
\def\BibTeX{{\rm B\kern-.05em{\sc i\kern-.025em b}\kern-.08em
    T\kern-.1667em\lower.7ex\hbox{E}\kern-.125emX}}

\usepackage[hidelinks,bookmarks=true]{hyperref}
\usepackage[nameinlink,capitalize]{cleveref}
\usepackage{makecell}

\newcommand{\REV}[1]{#1}
\newcommand{\ie}{i.e.,\xspace}
\newcommand{\eg}{e.g.,\xspace}
\newcommand{\etalcite}[1]{et al.~\cite{#1}}

\begin{document}

\title{Characterizing CPU-Induced Slowdowns in Multi-GPU LLM Inference}

\author{
\IEEEauthorblockN{
Euijun Chung,
Yuxiao Jia,
Aaron Jezghani,
Hyesoon Kim
}
\IEEEauthorblockA{
Georgia Institute of Technology
}
\IEEEauthorblockA{
\emph{
\href{mailto:euijun@gatech.edu}{\{euijun},
\href{mailto:yjia305@gatech.edu}{yjia305},
\href{mailto:ajezghani3@gatech.edu}{ajezghani3\}@gatech.edu},
\href{mailto:hyesoon@cc.gatech.edu}{hyesoon@cc.gatech.edu}
}}
}


\maketitle

\begin{abstract}
Large-scale machine learning workloads increasingly rely on multi-GPU systems, yet their performance is often limited by an overlooked component: the CPU.
Through a detailed study of modern large language model (LLM) serving workloads, we find that multi-GPU performance often degrades not because GPUs are saturated, but because CPUs fail to keep them busy. 
Under limited CPU allocations, systems exhibit symptoms such as delayed kernel launch, stalled communication, and increased tokenization latency, leading to severe GPU underutilization even when ample GPU resources are available. 
The problem becomes more severe in agentic LLM serving, where long accumulated contexts increase CPU-side tokenization work while high prefix-cache reuse across multi-turn interactions reduces GPU-side prefill work.
These bottlenecks persist even in serving stacks that employ process-level separation and modern GPU-side optimizations such as CUDA Graphs.
Since CPU cores cost orders of magnitude less than GPUs, provisioning additional cores is a highly cost-effective mitigation.
Under moderate serving load, we observe that CPU-starved configurations frequently time out, while providing adequate CPU resources restores responsiveness and reduces time-to-first-token (TTFT) latency by 1.47--7.11$\times$ across configurations, all without requiring additional GPUs.

\end{abstract}

\section{Introduction}

Modern machine learning (ML) workloads require massive computational power, driving the widespread adoption of multi-GPU servers, such as the DGX H100 and DGX B200~\cite{dgx-h100, dgx-b200}. GPU servers are expensive to purchase and operate, requiring substantial electricity and cooling infrastructure. To ensure efficient use of valuable resources, organizations operate them in a shared, multi-tenant manner, allowing multiple users to work on their own allocated virtual instances. Through resource schedulers such as Slurm or cloud platforms (\eg AWS and Azure), users request fixed, manually specified allocations (\eg 4 GPUs and 8 CPU cores) at a given cost per unit time~\cite{slurm, aws, weng2022mlaas}. While this approach improves GPU utilization across many users, it creates a critical blind spot: the CPU is often underprovisioned relative to GPUs. In this paper, we reveal why this imbalance can silently turn the CPU into a first-order bottleneck in multi-GPU large language model (LLM) inference, even when the GPUs themselves have ample compute capacity.

In multi-GPU ML workloads, CPUs perform numerous critical tasks, and systems with limited CPU resources can experience severe slowdowns when CPUs are oversubscribed. 
The CPU's tasks include data I/O and processing, inter-GPU synchronization, request scheduling, and kernel launches~\cite{nccl, li2020pytorch, liu2024deepseek}. 
A significant portion of a CPU's job involves feeding sufficient data to GPUs, \ie retrieving data from host memory and performing preprocessing steps such as tokenization for language models~\cite{huggingface-tokenizers}, or image decoding and augmentation for vision models~\cite{krizhevsky2012imagenet, DALI, wu2020visual, touvron2021training-deit}.
Moreover, CPUs are responsible for coordinating computation and communication across multiple GPUs by issuing kernel launches, managing stream dependencies, and orchestrating synchronization~\cite{kwon2023efficient, wong2025gpus, chang2026llm}.

\begin{figure}[t]
    \centering
    \includegraphics[width=0.9\linewidth]{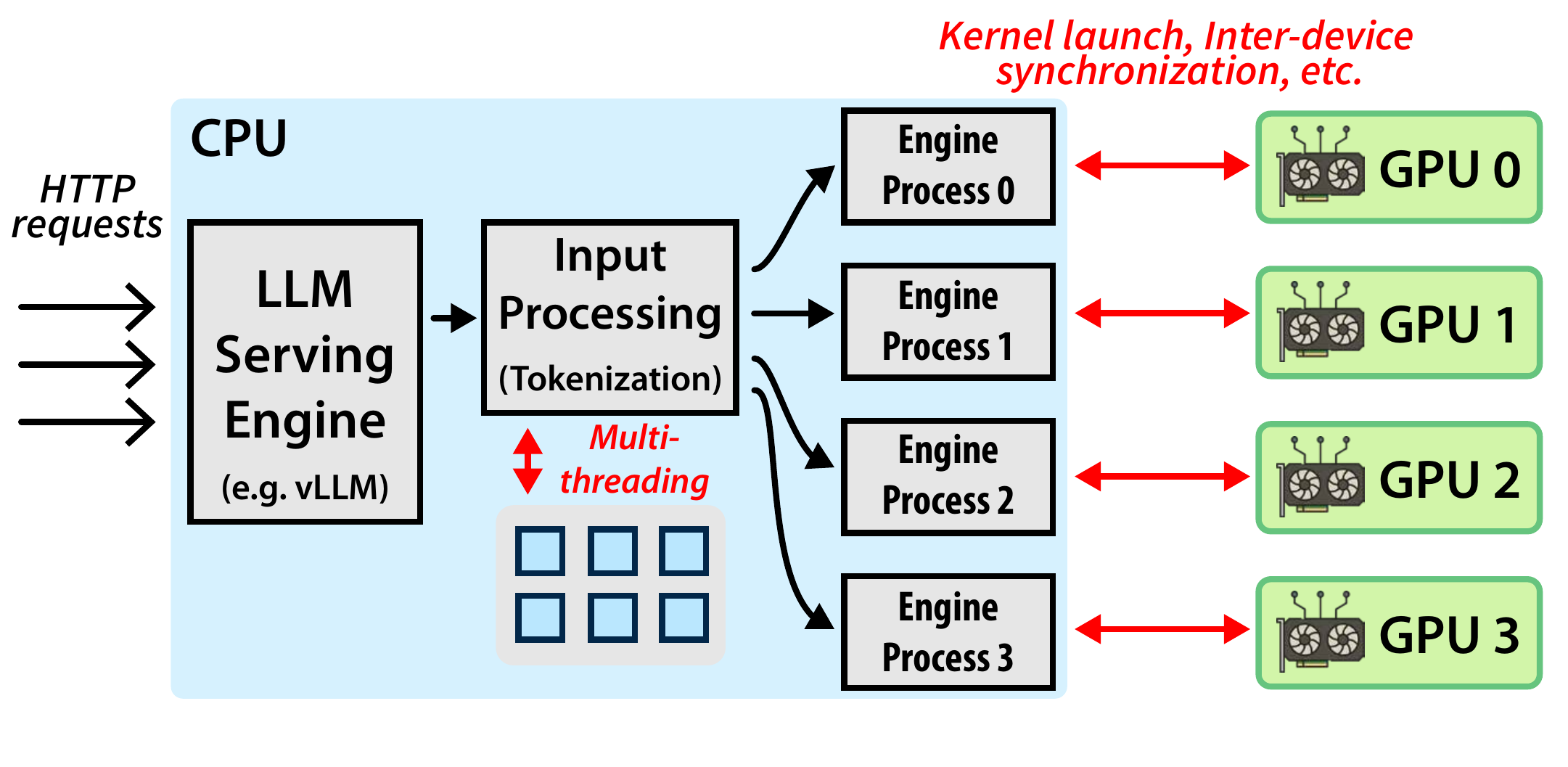}
    \vspace{-0.15in}
    \caption{CPU tasks in multi-GPU LLM serving: input processing, kernel launches, and synchronization.\label{fig:summary}}
    \vspace{-0.1in}
\end{figure}

However, once such processes contend for limited CPU cores, oversubscription increases host-side latency and slows the launch of both compute and communication kernels. This behavior is especially pronounced during LLM serving with multiple incoming requests, where a high CPU load from these requests can substantially increase kernel launch latency from microseconds to milliseconds, thereby degrading overall performance and risking service-level objective (SLO) violations~\cite{zhao2025insights, gujarati2020serving}.

Collective communication libraries, such as NCCL~\cite{nccl, hu2025demystifying}, can lead to additional slowdowns in scenarios with limited CPU resources, as their collectives require all participating GPUs to synchronize. If the CPU is late to dispatch on any rank, the entire device group stalls. While CUDA Graphs~\cite{cudagraph} amortize launch overhead for static kernel sequences, they cannot capture dynamic scheduling decisions required at every decode step of LLM inference~\cite{durvasula2024acs}, leaving CPU-side bottlenecks as a persistent limiting factor.

The CPU bottleneck becomes even more severe in agentic and multi-turn LLM serving. 
As contexts grow across turns, high prefix-cache reuse reduces GPU-side prefill work, while the CPU must still tokenize the full accumulated context at every turn. 
This imbalance increases the relative importance of CPU-side work and makes the system more sensitive to CPU scarcity.

The goal of this work is to challenge a common misconception among multi-GPU system users: the assumption that GPU performance alone determines the efficiency of modern ML workloads. Prior studies have shown that host-side orchestration and CPU single-thread performance can affect LLM inference latency in single-GPU or CPU-GPU coupled settings~\cite{vellaisamy2025characterizing, vellaisamy2026taxbreak}, and practitioners have noted that CPU performance matters in multi-GPU systems~\cite{scaling, zhao2025insights, rising2026varra}. However, it remains unclear why limited CPU resources can cause severe slowdowns, specifically in multi-GPU LLM serving, where the symptoms often appear as low GPU utilization, time-to-first-token (TTFT) spikes, or request timeouts.

This paper explains why CPU scarcity becomes a performance bottleneck in multi-GPU LLM serving by identifying three critical host-side paths. First, tokenization threads compete with latency-sensitive LLM engine processes for CPU cores, delaying scheduling and kernel dispatch. Second, late CPU-side kernel launches on any GPU rank create stragglers at collective synchronization points, forcing other GPUs to wait. Third, the shared-memory broadcast queue used to deliver scheduling metadata from the engine to GPU workers can itself become a CPU-side delay on the critical path. Together, these mechanisms explain how CPU-side delays propagate into GPU idle time and end-to-end latency, even in optimized serving stacks with CUDA Graphs, \texttt{torch.compile}, and process-level isolation.

We evaluate these effects under varying CPU resource configurations in LLM inference and online serving scenarios in \Cref{sec:inference}, then analyze the underlying synchronization and shared-memory bottlenecks in \Cref{sec:understanding}. \Cref{sec:discussion} discusses deployment implications, including a case study on a high-performance computing (HPC) cluster and cloud CPU economics. \Cref{sec:related} reviews related work, and \Cref{sec:conclusion} concludes.

Our contributions are as follows:

\begin{itemize}
\item \textbf{Characterization.} In long-context LLM serving with high prefix-cache reuse, we show that CPU scarcity can substantially increase TTFT. Across dense and MoE models on 4--8 GPUs, CPU-starved configurations experience up to 7.11$\times$ higher TTFT and engine timeouts compared with CPU-abundant configurations.
\item \textbf{Mechanisms.} We identify two mechanisms by which CPU contention causes end-to-end slowdowns: delayed kernel dispatch interacting with collective synchronization barriers, which inflates collective time by 3.47$\times$, and tail stalls in the engine-to-worker broadcast path that delay GPU workers.
\item \textbf{Deployment relevance.} We analyze 4.65 million real-world cluster allocation records to show that CPU underprovisioning is common in practice, and demonstrate that increasing CPU allocation can be a practical mitigation given its relatively low marginal cost compared with GPU resources.
\end{itemize}

\section{Background \& Motivation}
\subsection{CPU's Role in LLM Serving}

In modern LLM serving systems, the CPU is responsible for several latency-sensitive operations that directly influence end-to-end performance. As shown in \Cref{fig:summary}, the CPU's job in LLM inference includes: \textbf{\textcircled{\scriptsize{1}} processing and tokenizing input text}, \textbf{\textcircled{\scriptsize{2}} handling HTTP requests and scheduling them}, and \textbf{\textcircled{\scriptsize{3}} issuing GPU kernel launches} to sustain steady execution on the accelerators. Although GPUs perform the bulk of the model's numerical computation, the CPU remains responsible for coordinating each step of computation, and these input processing and control-plane tasks can become the dominant bottleneck in multi-GPU inference.

\begin{figure}[t]
    \centering
    \includegraphics[width=0.9\linewidth]{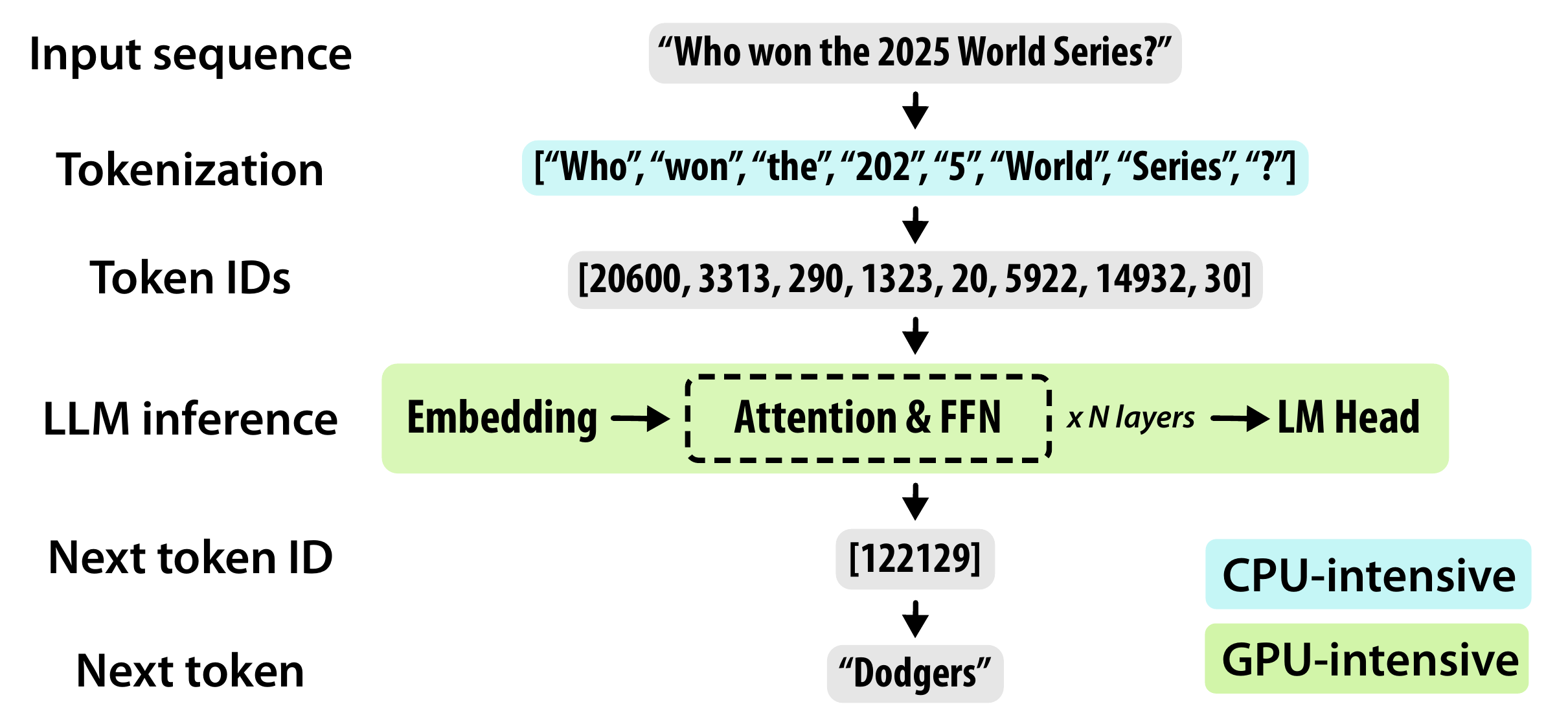}
    \vspace{-0.1in}
    \caption{LLM inference pipeline: CPU-intensive tokenization followed by GPU-intensive forward pass of the LLM.\label{fig:tokenization}}
    \vspace{-0.1in}
\end{figure}

\textbf{\textcircled{\scriptsize{1}} Tokenization} adds substantial CPU work during LLM inference. As shown in \Cref{fig:tokenization}, tokenization converts raw text strings into integer token IDs that the model can process using subword segmentation algorithms such as byte pair encoding (BPE)~\cite{sennrich2016neural} or SentencePiece~\cite{kudo2018sentencepiece}. This process occurs at the beginning of every inference request for prompt processing. Tokenization consumes substantial CPU cycles, particularly for long prompts or batched requests, where preprocessing costs scale linearly with the input size. Because tokenization is a prerequisite to any LLM computation, it lies on the critical path of each request and adds directly to end-to-end latency unless overlapped with the GPU-side computation of other requests.

Widely used tokenizer implementations employ multiprocessing or multithreading to spread the overhead of text processing across multiple CPU cores. In particular, the HuggingFace Tokenizers library~\cite{huggingface-tokenizers} enables its Rust-based tokenizer to spawn multiple parallel threads, with \texttt{TOKENIZERS\_PARALLELISM=true} as the default setting.

\textbf{\textcircled{\scriptsize{2}} HTTP server request handling} also adds CPU load through connection management, request parsing, and batched scheduling. This overhead is typically much smaller than the tokenization cost, so our analysis focuses primarily on tokenization.

\begin{table*}[t]
    \centering
    \caption{List of CPU-GPU heterogeneous system setups used in this paper's evaluation.\label{tab:system-setup}}
    \vspace{-0.05in}
    \begin{tabular}{c || c l c c l} 
        \Xhline{2\arrayrulewidth}
        \textbf{System (GPU)} & \textbf{\makecell{Architecture \\ (Compute Capability)}} & \textbf{CPU Model} & \textbf{\#CPU Cores} & \textbf{\#GPUs per Node} & \textbf{Inter-GPU Interconnect} \\
        \Xhline{2\arrayrulewidth}
        H100 & Hopper (9.0) & Intel Xeon Platinum 8462Y+ & 64 & 8 & NVLink 4.0 (900 GB/s bidirectional) \\

        H200 & Hopper (9.0) & Intel Xeon Platinum 8562Y+ & 64 & 8 & NVLink 4.0 (900 GB/s bidirectional) \\
        
        RTX Pro 6000 & Blackwell (12.0) & Dual Intel Xeon 6737P & 64 & 8 & No NVLink (PCIe 5.0, 64 GB/s) \\
        
        \Xhline{2\arrayrulewidth}
    \end{tabular}
    \vspace{-0.1in}
\end{table*}

\textbf{\textcircled{\scriptsize{3}} Kernel launch overhead} represents another source of CPU work. The CPU must schedule and dispatch CUDA kernels to sustain model execution, intervening at least once per generated token: each token requires host-side processing, such as sampling, end-of-sequence (EOS) detection, or agentic tool-call dispatch, before the next step can be launched. If the host thread responsible for kernel dispatch is delayed by OS scheduling or CPU contention, the GPU idles until the launch completes, slowing the entire pipeline.

CUDA Graphs~\cite{cudagraph} amortize most repeated launch calls by capturing and replaying complete computation sequences. However, they cannot capture operations that remain dynamic, including stream synchronizations, library kernels outside the graph, and the per-token host-side processing above. CUDA graph launch must still be issued by the CPU at every decode step~\cite{durvasula2024acs, kim2025cost}.

To ensure that GPUs receive kernels on time, multi-GPU ML and LLM libraries such as \texttt{torch.distributed}~\cite{li2020pytorch} rely on multiprocessing to allow each process to launch kernels independently and provide sufficient concurrency for high-throughput GPU dispatch. As a result, multi-GPU inference frameworks, including vLLM~\cite{kwon2023efficient}, DeepSpeed~\cite{rasley2020deepspeed}, and HuggingFace Accelerate~\cite{accelerate} allocate at least one process per GPU, as also depicted in \Cref{fig:summary}. The key takeaway is that modern multi-GPU frameworks should assign at least one process per GPU to enable efficient kernel dispatch and consistently keep the accelerators fed.

Taken together, when CPU cores are underprovisioned, data-processing tasks compete with the LLM engine processes for cores, and the entire GPU workload pipeline can be substantially delayed. 
In particular, under concurrent LLM requests, multiple requests can keep the tokenizer heavily occupied and contend with the kernel-launching processes, introducing delays even on systems with seemingly sufficient CPU resources. 
This competition causes cascading slowdowns in kernel dispatch and request scheduling, ultimately degrading the end-to-end performance. 
We present experiments in \Cref{sec:inference} that show how the CPU can cause end-to-end slowdowns, along with an analysis of latency and CPU and GPU utilization. 
Additionally, \Cref{sec:communication-microbenchmark} examines the effect of CPU oversubscription in more detail, visualizing cases in which CPU-side resource contention delays kernel calls and results in GPU underutilization.

\REV{\subsection{Agentic LLM Workloads and Prefix Caching}
With the rise of agentic AI, LLM inference increasingly runs inside multi-turn loops that interleave reasoning with tool invocations~\cite{kim2025cost, raj2025cpu, yao2022react, zhang2026toktier}.
Agentic and multi-turn workloads resubmit the entire accumulated context at every turn: the model re-reads the conversation history plus a short appended delta, such as a tool result, before producing its next action~\cite{liu2026agentic, li2025continuum, yu2025stateful}. 
Trace studies show that while contexts accumulate to tens or hundreds of thousands of tokens over a session, each turn appends only a few hundred new tokens~\cite{ding2026observation, qin2024mooncake}. Serving engines therefore rely on prefix caching to reuse the key-value (KV) cache of previously processed tokens; in such sessions, over 95\% of each turn's input tokens are served from the cache, reducing per-turn GPU prefill to the small uncached delta~\cite{gao2024cost, wu2026dualpath, yuan2026kairos, zhu2026tracelab, gim2024prompt}.

Prefix caching, however, does not reduce the CPU-side cost of a turn. In mainstream engines such as vLLM~\cite{kwon2023efficient} and SGLang~\cite{zheng2024sglang}, a request must be fully tokenized before the cache can be queried, so each turn re-tokenizes the entire accumulated context. This cost grows with session length~\cite{shao2025lopt}, while the GPU prefill remains bounded by the small per-turn delta. As cache hit rates rise, the CPU accounts for an increasing share of TTFT, and this shift is the regime our work characterizes.}

\section{Multi-GPU System Evaluation Setup}
\label{sec:setup}

\Cref{tab:system-setup} lists the H100, H200, and RTX Pro 6000 systems used in our evaluation. We disable simultaneous multithreading (SMT) to isolate effects from physical cores. Since cloud platforms typically expose vCPUs, our measurements provide a conservative baseline for CPU provisioning. We run our evaluations on the listed machines with restricted (affinity-limited) CPU allocations to examine how multi-GPU setups in multi-tenant clusters can lead to unexpected performance degradation due to the limited CPU resources. For example, on a server equipped with four H100 GPUs, we may create a constrained execution environment with only 16 CPU cores to emulate a realistic cloud or cluster.

All LLM experiments use vLLM v0.21.0 with the V1 engine, which separates API serving/tokenization from EngineCore scheduling and GPU worker execution across processes. Default optimizations, including CUDA Graphs, prefix caching, \texttt{torch.compile}, \REV{asynchronous scheduling}, and custom \texttt{AllReduce} kernels, remain enabled. Thus, the observed bottlenecks persist in an optimized serving stack rather than an intentionally unoptimized baseline.

Moreover, we use a fixed CPU \& GPU resource allocation per job, reflecting common practice in cluster schedulers such as Slurm~\cite{slurm} and in cloud service offerings where users receive a predetermined number of CPU cores per GPU~\cite{aws}. While such isolation supports fair sharing across users, it also introduces the risk of mismatched CPU and GPU allocations, as these allocations are typically chosen manually or determined by simple heuristics (\eg assigning 4 CPU cores per GPU).

\section{The CPU Bottleneck in LLM Inference}

\label{sec:inference}

In this section, we analyze how CPU-side tasks during LLM inference and serving, such as tokenization, runtime orchestration, and communication coordination, can become dominant bottlenecks when multiple jobs or requests execute concurrently. We show that contention for limited CPU resources not only delays host-side processing but also slows GPU execution, resulting in increased per-token latency and reduced overall throughput.

\subsection{Tokenization Latency Evaluation in LLM Inference}

\begin{figure}[t]
    \centering
    \includegraphics[width=0.9\linewidth]{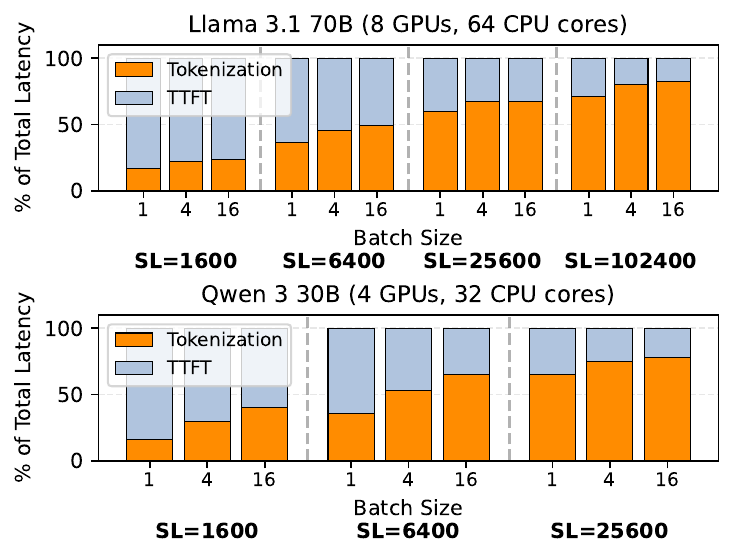}
    \vspace{-0.15in}
    \caption{\REV{Tokenization vs. time-to-first-token (TTFT) latency across batch sizes and sequence lengths (SL). Llama on 8$\times$H200 (64 CPU cores). Qwen on 4$\times$H200 (32 CPU cores).\label{fig:token-eval}}}
    \vspace{-0.1in}
\end{figure}

\Cref{fig:token-eval} visualizes how much of the end-to-end LLM prefill latency is attributable to tokenization. We use vLLM~\cite{kwon2023efficient} to systematically quantify the relative cost of CPU-side tokenization compared to GPU-side model execution; similar CPU-side bottlenecks are expected in other frameworks such as SGLang~\cite{zheng2024sglang} with a multi-process architecture. We evaluate two models with distinct sparsity characteristics: a dense model (Llama 3.1 70B) on an 8$\times$H200 system with \REV{64 CPU cores}, and a Mixture-of-Experts (MoE) model (Qwen3 30B, 3B active parameters) on a 4$\times$H200 system with \REV{32 CPU cores}. Both use tensor parallelism (TP)~\cite{shoeybi2019megatron} and HuggingFace Tokenizers~\cite{huggingface-tokenizers} as integrated in vLLM. 
We measure TTFT as the wall-clock duration of \texttt{llm.generate()} function with pre-tokenized inputs (engine-internal tokenization bypassed), averaged over five iterations after three warmup runs.
We measure tokenization latency separately via \texttt{tokenizer.encode()} calls. 

\REV{For each trial, we vary batch size $\{1, 4, 16\}$ and input sequence length $\{1600, 6400, 25600, 102400\}$, ranging from a single interactive request to a moderately batched, long-context multi-tenant load~\cite{qin2024mooncake}. To emulate steady-state agentic turns, each request appends a fresh 256-token tool-call result to a context with a warmed prefix cache, reflecting observed per-turn appends of a few hundred tokens~\cite{ding2026observation}.}

\noindent \textbf{Results.} Tokenization accounts for up to roughly 80\% of TTFT in long-sequence, large-batch configurations across both dense (Llama) and MoE (Qwen) architectures. The fraction grows with sequence length by construction of this regime. Tokenization processes the full accumulated context, whereas GPU prefill computes only the fixed 256-token delta; the CPU-side cost therefore scales with sequence length while the GPU-side cost stays nearly constant. \REV{As sessions approach million-token contexts, per-turn tokenization alone can require seconds of CPU time~\cite{wang2024beyond}. While the relative impact of tokenization decreases once decode is taken into account, agentic sessions re-tokenize the full context at every turn, so tokenization returns to the critical path at each turn's TTFT.}

\subsection{Impact of Tokenization Load in LLM Serving}

Next, we evaluate how a busy tokenizer can slow down the LLM engine. Our experiment shows that under heavy load, multiple tokenizer threads can compete with the LLM engine processes for CPU resources, which leads to frequent context switches, delayed kernel launches, and reduced GPU utilization. Such a slowdown substantially degrades latency as communication kernels rely on barrier-based synchronization, as discussed further in \Cref{sec:communication-microbenchmark}.

\begin{figure}[t]
    \centering
    \includegraphics[width=1.0\linewidth]{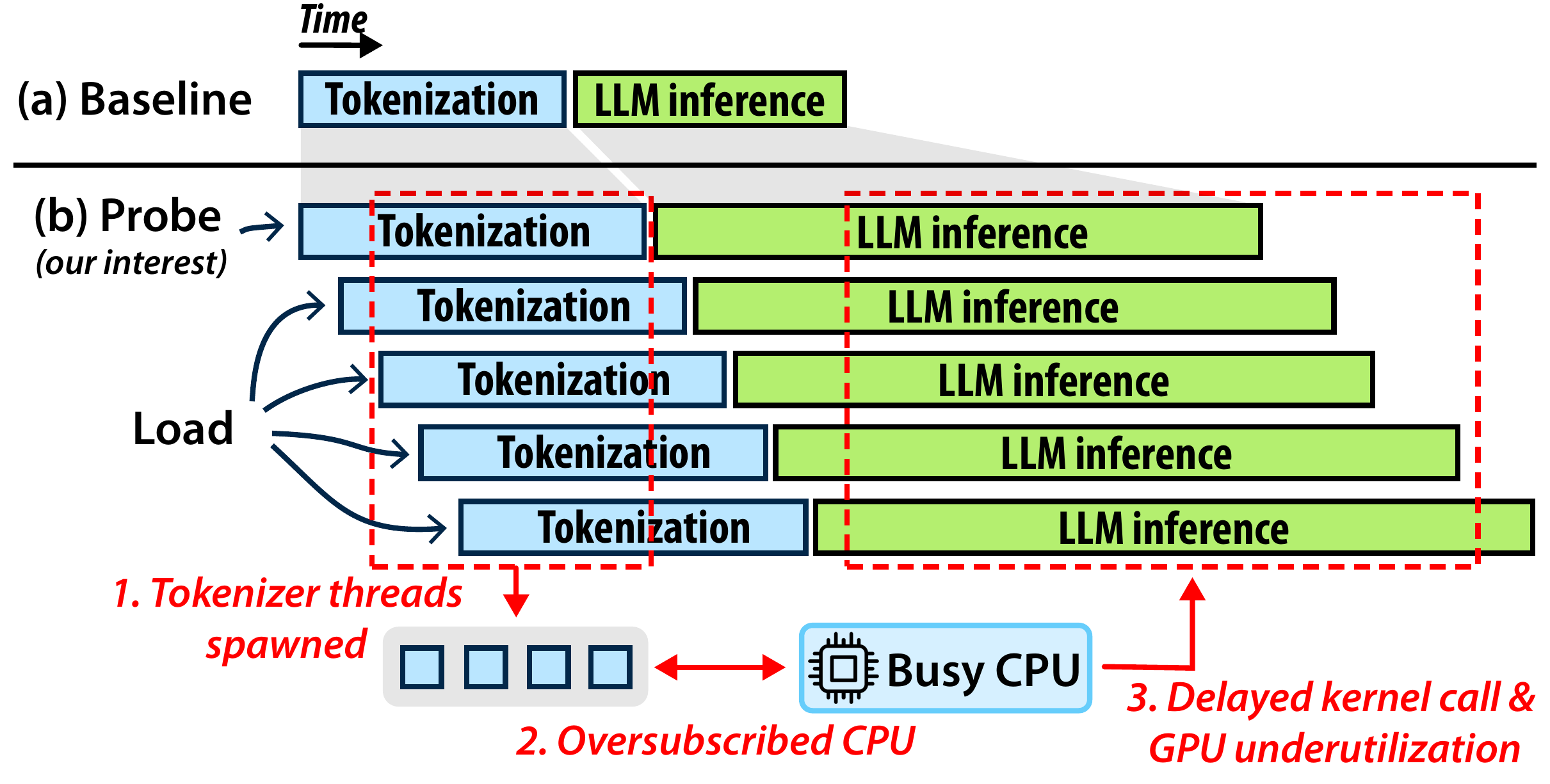}
    \vspace{-0.25in}
    \caption{(a) Baseline LLM inference. (b) \REV{Probe under load}: (1) tokenizer spawns threads, (2) CPU saturates, and (3) delayed kernel launches underutilize GPUs.\label{fig:attacker}}
    \vspace{-0.1in}
\end{figure}

\begin{figure*}[t]
    \centering
    \includegraphics[width=0.95\linewidth]{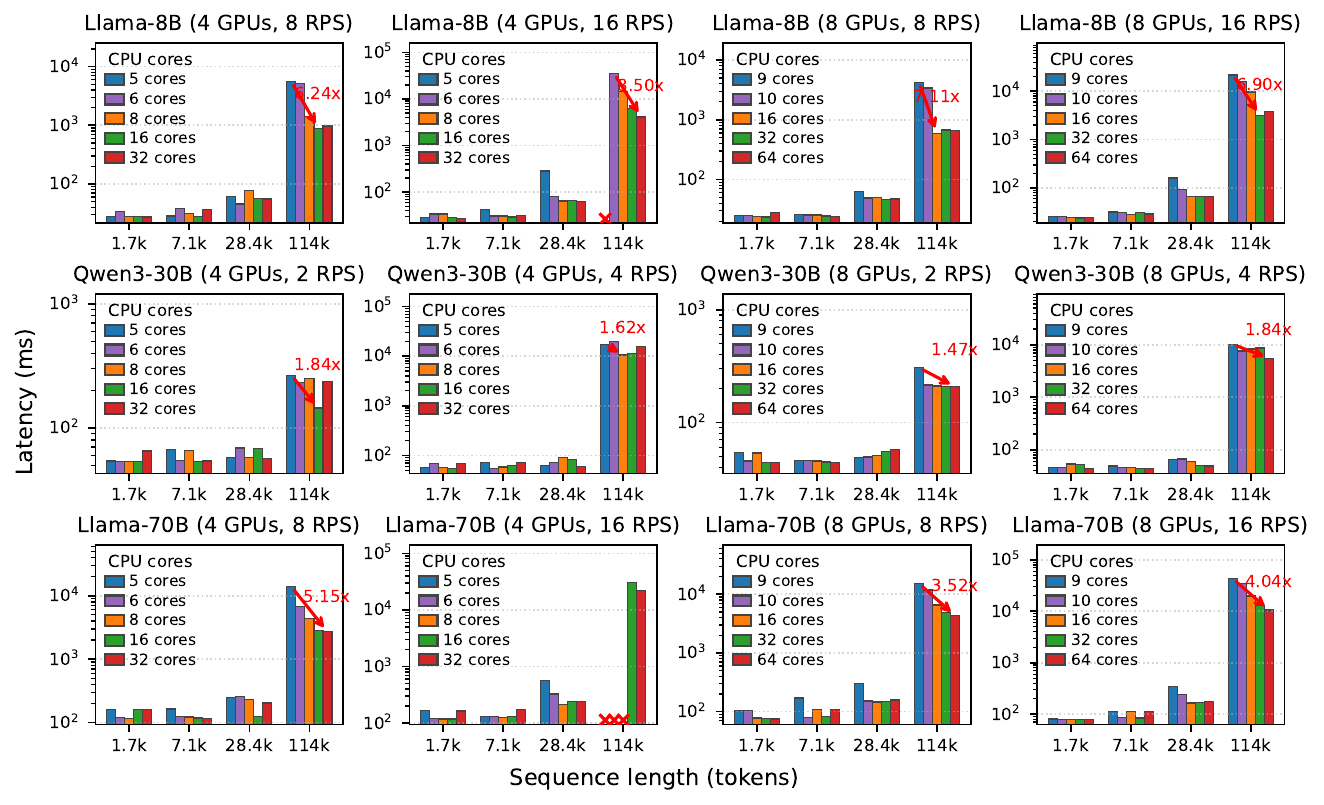}
    \vspace{-0.15in}
    \caption{\REV{Probe TTFT across models on 4- and 8-GPU configurations: Llama 3.1 8B (H100, 8/16 requests per second (RPS)), Qwen3 30B (H100, 2/4 RPS), Llama 3.1 70B (H100, 8/16 RPS). Red $\times$ marks 200-second timeouts; red arrows show speedup from smallest to best CPU allocation.\label{fig:attack-results}}}
    \vspace{-0.1in}
\end{figure*}

\noindent \textbf{Evaluation methodology.} We design an experiment where multiple concurrent requests are sent to the LLM server, and we measure how this background load affects the latency of a single target request. We refer to the measured request as the \textit{probe}, and we generate additional periodic requests, termed \textit{load}, that impose a controlled CPU load defined by a specified requests per second (RPS) rate. 
\REV{Note that both request types are served by a single LLM engine and batched together through continuous batching, sharing the same model instance without any partitioning of CPU or GPU resources~\cite{yu2022orca}.}

The overall setup is illustrated in \Cref{fig:attacker}, which shows how a four-request load after the probe keeps the tokenizer busy, thereby delaying both tokenization and inference due to delayed kernel calls and GPU underutilization. We implement this experiment using \texttt{vllm serve} with the V1 async engine described in \Cref{sec:setup}, where user queries arrive asynchronously over HTTP. Despite V1's process-level separation of tokenization, scheduling, and GPU execution, CPU contention persists because all processes compete for the same limited pool of CPU cores. In this experiment, we quantify CPU contention by measuring the latency of the probe under load while varying the number of CPU cores. This approach distinguishes CPU- versus GPU-limited regimes in multi-GPU LLM serving.

\noindent \textbf{LLM serving system setup.} We evaluate three models spanning sizes and sparsity regimes: Llama 3.1 8B (dense), Qwen3 30B (MoE), and Llama 3.1 70B (dense), each in 4-GPU (TP=4) and 8-GPU (TP=8) configurations. All models run in bfloat16 (bf16) precision. CPU resources (\#CPU) are provisioned at five levels: (\#GPUs + 1) cores (least-CPU case), (\#GPUs + 2) cores (matching vLLM V1's concurrent-process count), 2$\times$\#GPUs, 4$\times$\#GPUs, and 8$\times$\#GPUs (CPU-abundant cases). As described in \Cref{sec:setup}, vLLM V1 requires at least (\#GPUs + 2) concurrent processes; our least-CPU configuration of (\#GPUs + 1) cores is therefore already below this minimum, guaranteeing that at least one process must time-share a core via OS scheduling. In practice, allocating exactly \#GPUs cores causes the system to operate prohibitively slowly, so we use (\#GPUs + 1) as the bare-minimum functional configuration. For every configuration, we report the client-side TTFT under the assumption of a prefill-decode-disaggregated deployment for the prefill stage~\cite{zhong2024distserve, patel2024splitwise, qin2024mooncake}.

The load's RPS values differ across models: Llama uses 8 and 16, while Qwen3 30B uses 2 and 4 because the higher rates time out across all cells for the sparse MoE Qwen3 30B.
The load's sequence length ranges from 1.8k to 114k tokens, while the probe's sequence length is 2.8k tokens. There are slight differences in the exact token count between the Llama and Qwen models. We measure the average latency of four probes after dropping the first post-load sample as a warmup. 
We set the output length of each load request to one token to isolate long-context request processing from sustained decode load.
\REV{Both parameters reflect realistic serving conditions~\cite{xiang2026servegen, wang2024burstgpt}: the request rates fall within observed production ranges (the tool-and-agent workload trace in Mooncake~\cite{qin2024mooncake} averages 6.7 RPS, with bursts up to 23).
The sequence lengths also represent moderate load for long-context serving, given that recent models such as Qwen3.5-Flash~\cite{team2026qwen3} and Nemotron~3~\cite{blakeman2025nvidia} support context windows of up to 1M tokens.
Prefix caching is enabled, reflecting reuse from shared system prompts and accumulated tool-call history for the load. Our synthetic load reaches KV cache hit rates of 92.5\%, 78.1\%, and 95.3\% for Llama 8B, Qwen 30B, and Llama 70B, respectively. However, each probe carries a unique prefix and is served entirely uncached, so its  TTFT includes tokenization and a full prefill pass.}

\noindent \textbf{Latency results.} \Cref{fig:attack-results} summarizes the TTFT of probes under a load at a given RPS. Across the configurations, increasing \#CPU can eliminate engine timeouts (marked in red $\times$) or reduce the probe's TTFT, confirming that the CPU bottleneck is not specific to a single system or a model. Removing CPU scarcity improves long-sequence TTFT by 1.47--7.11$\times$ when we go from (\#GPUs + 1) cores to a CPU-abundant configuration.

This improvement stems from two effects: \textcircled{\scriptsize{1}} tokenizer threads in the API server face less contention and finish sooner, and \textcircled{\scriptsize{2}} host-side tasks across the EngineCore (batching, scheduling, and memory management) and GPU worker processes (kernel dispatch and data transfers) proceed without delay. \REV{Even at (\#GPUs + 2) cores, where every engine process holds a dedicated core, the probe's TTFT remains 2.63$\times$ (geometric mean) above the CPU-abundant configuration: the tokenizer alone spawns more threads than the remaining cores, so contention persists until CPU cores well outnumber the engine's processes and threads.}

With higher load RPS, fewer-core systems saturate earlier: context switching spikes, kernel launches serialize, and GPUs sit idle even when work is available. Larger CPU allocations tolerate much higher load before throughput collapses, so CPU headroom directly raises the sustainable request rate. However, at the largest 64-core allocation, we occasionally observe a slight latency rebound, likely from cross-socket non-uniform memory access (NUMA) effects and over-parallelization in the tokenizer's thread pool.

\noindent \textbf{MoE models are more sensitive to CPU scarcity than dense models.} In our load experiments, Qwen3 30B (MoE) reaches engine saturation at far lower rates (RPS=2--4) than Llama 3.1 70B (dense) on the same H100 system (RPS=8--16). 
Because an MoE model activates only a few experts per request, its per-step GPU time is shorter, so CPU-side work occupies a larger fraction than in a dense model~\cite{fedus2022switch}.
This empirically corroborates the single-node host-bound analysis in TaxBreak~\cite{vellaisamy2026taxbreak}, which reports that MoE inference is particularly sensitive to host CPU performance. Our multi-GPU concurrent-serving experiments extend that observation: under sustained load, the smaller per-step GPU usage of MoE amplifies CPU-side queueing and produces an earlier transition to the CPU-bound regime.

\begin{figure}[t]
    \centering
    \includegraphics[width=0.9\linewidth]{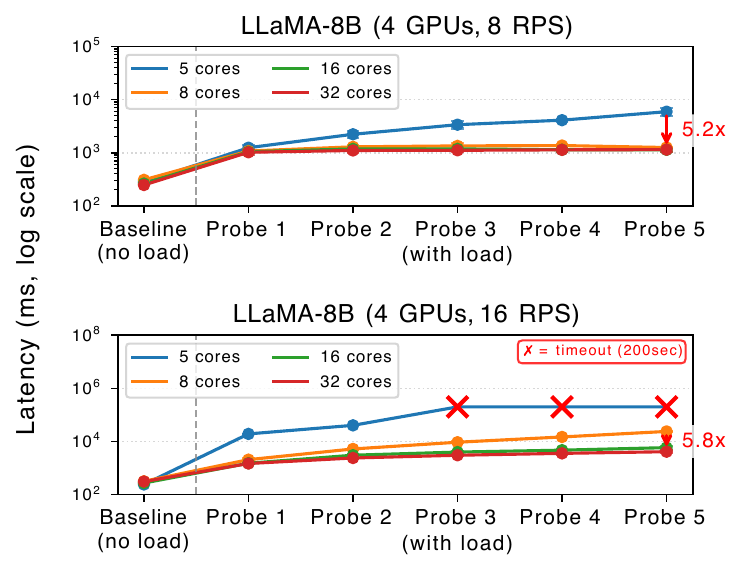}
    \vspace{-0.15in}
    \caption{Sequentially issued \REV{probes'} TTFT under load (8/16 requests per second (RPS), 114k tokens); more CPU cores reduce the slowdown. Measured on Blackwell GPUs.\label{fig:attack-per-request}}
    \vspace{-0.1in}
\end{figure}

\noindent \textbf{LLM engine timeouts.} Some configurations fail to produce latency values, marked by red $\times$ in \Cref{fig:attack-results}. The failure follows a positive-feedback loop. Tokenization from the load saturates the CPU, delaying scheduling and kernel launches. Then, the resulting higher latency keeps requests active longer, so more arrivals accumulate in the engine queue and further amplify tokenization and scheduling contention. Eventually, the system enters an overloaded state where the request queue grows unbounded and the probe request times out at our 200-second limit~\cite{nie2026queueing}. The feedback is accelerated on MoE models: if concurrent requests activate different experts, the batch touches more expert weights per step, lengthening per-step GPU time and growing the queue at far lower request rates.

\noindent \textbf{Latency per request.} \Cref{fig:attack-per-request} reports the TTFT latencies of the baseline and five probes under load of 8/16 RPS on Llama 3.1 8B with TP=4, sequence length (SL)=114k. Probes are issued sequentially, so as the load's requests accumulate in the serving engine, each subsequent probe's TTFT shows an increasing trend. In particular, scaling from 5 CPU cores (the least-CPU configuration) to 32 cores consistently yields over 5$\times$ lower TTFT.

\begin{figure}[t]
    \centering
    \includegraphics[width=0.8\linewidth]{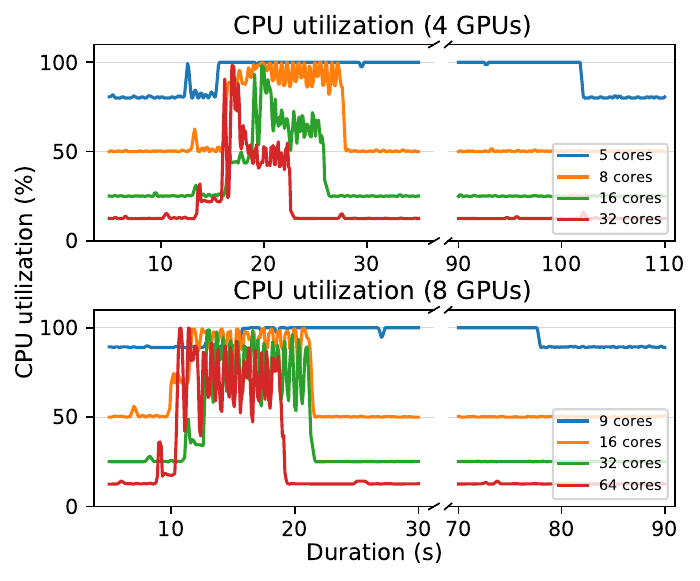}
    \vspace{-0.15in}
    \caption{CPU utilization (Llama 3.1 8B) across core allocations on 4-/8-GPU configurations; more cores shorten the saturation period.\label{fig:cpu-utilization-4v8-gpu}}
    \vspace{-0.1in}
\end{figure}

\begin{figure}[t]
    \centering
    \includegraphics[width=0.8\linewidth]{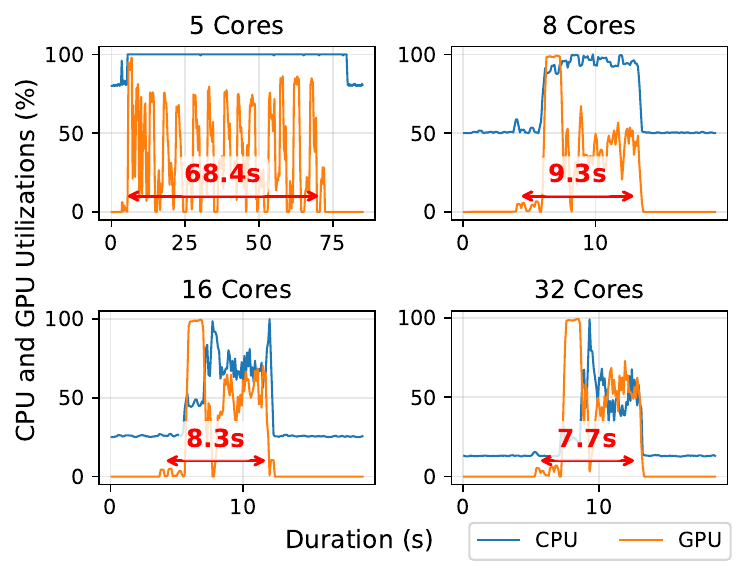}
    \vspace{-0.15in}
    \caption{CPU/GPU utilization (Llama 3.1 8B, 4-GPU) across CPU allocations: CPU saturation leaves GPUs idle.\label{fig:cpu-gpu-utilization}}
    \vspace{-0.1in}
\end{figure}

\noindent \textbf{CPU and GPU utilization.} \REV{We sample CPU and GPU utilization at 100\,ms intervals using \texttt{psutil} and \texttt{nvidia-smi}, under 8~RPS with 114k-token sequences.} 
Traces are shown for Llama only, as Qwen3 30B behaves qualitatively similarly. 
\Cref{fig:cpu-utilization-4v8-gpu} shows that CPU-induced latency correlates more strongly with the duration of CPU saturation than with peak utilization.
With 5 cores, the CPU stays near 100\% for up to 100 seconds, and preprocessing tasks queue behind it. With 8--32 cores, utilization still spikes, but the spikes drain within seconds, and the GPU launch path remains supplied with work; the 8-GPU configuration behaves the same up to 64 cores. \Cref{fig:cpu-gpu-utilization} shows the consequence on the GPU side: during CPU saturation, the GPU sits idle, and with sufficient cores the same work finishes in a visibly shorter span. \REV{Note that \texttt{nvidia-smi} counts a GPU as busy even while kernels spin-wait at synchronization barriers, so these traces measure occupancy rather than useful work; \Cref{sec:kernel-profile} quantifies effective kernel execution directly.}

\REV{\noindent \textbf{Effect of further host-side optimizations.}
We argue that while host-side optimizations continue to emerge, the CPU bottleneck is still likely to cause slowdowns and engine timeouts. 
For instance, \texttt{fastokens}~\cite{fastokens} is a recent tokenizer backend that outperforms HuggingFace Tokenizers on long contexts. 
To quantify the remaining CPU overhead, we repeat the serving experiment for Llama~3.1~70B (4 GPUs) with \texttt{fastokens} enabled. The faster tokenizer only raises the load threshold at which the engine collapses: the least-CPU configuration, which previously timed out at 16~RPS, now completes. However, its TTFT remains 2.13$\times$ higher than the CPU-abundant configuration, and the engine still times out at 28~RPS. Host-side optimizations thus shift the onset of CPU saturation to higher load but do not remove the CPU from the critical path.}

\noindent \textbf{Takeaways.} In multi-request LLM serving pipelines, tokenization adds substantial pressure on the CPU, and this often causes end-to-end latency to be limited not by GPU throughput but by the CPU's ability to perform tokenization and kernel scheduling on time. Under concurrent load, these CPU bottlenecks delay kernel launches, stall preprocessing pipelines, and leave GPUs underutilized. The ``probe under load'' experiment shows that allocating sufficient CPU resources is crucial for achieving high serving performance and preventing engine timeouts, particularly under high-RPS or long-context workloads where tokenization dominates the total work.

\section{Understanding the CPU Bottleneck in Multi-GPU Systems}
\label{sec:understanding}

This section analyzes how exactly CPU oversubscription and resource contention among multiple processes degrade multi-GPU performance by examining the CPU's role in kernel launches and GPU synchronization. We show that since communication kernels such as \texttt{AllReduce} synchronize all GPUs at a barrier, every GPU must busy-wait for the slowest rank, and this can stall the entire group when CPU resources are insufficient. Additionally, we find that contention on shared regions such as the Linux shared-memory segment increases launch-path latency when many processes compete for access. 

\begin{figure}[t]
    \centering
    \includegraphics[width=1.0\linewidth]{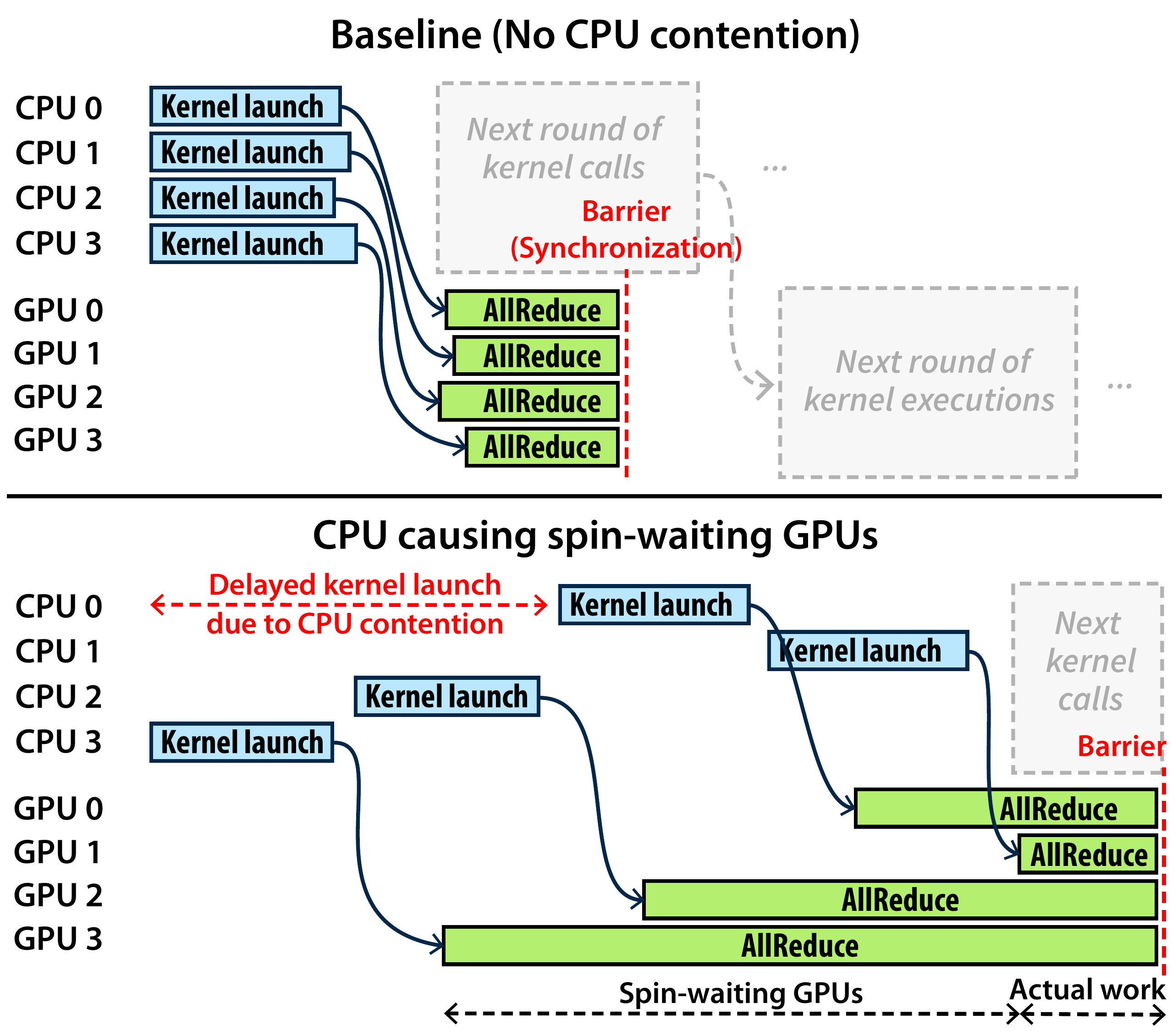}
    \vspace{-0.15in}
    \caption{Profiled timeline showing how contended CPU cores serialize kernel launches and underutilize GPUs. Gray boxes mark subsequent kernel calls.\label{fig:inference}}
    \vspace{-0.1in}
\end{figure}

\subsection{Synchronization and CPU Oversubscription in Communication Kernels}
\label{sec:communication-microbenchmark}

To examine how CPU contention affects GPU utilization, we develop a microbenchmark with the \texttt{torch.distributed} framework~\cite{li2020pytorch} and profile it using the PyTorch Profiler~\cite{pytorch-profiler}. The experiment isolates CPU-side kernel dispatch and GPU-side execution, allowing us to quantify the synchronization overheads induced by CPU oversubscription.

\Cref{fig:inference} illustrates the profiled result of communication kernels under extreme CPU contention, where multiple processes contend for a single CPU core. In this configuration, the CPU can launch only a single kernel at a time, forcing the GPU kernels to start sequentially rather than concurrently. Since NCCL collectives such as \texttt{AllReduce} introduce global synchronization barriers, faster GPUs must wait until every process has reached the barrier, creating idle busy-wait periods that waste compute cycles (highlighted in the figure by black dotted lines). This produces a \textit{straggler} effect: in an 8-GPU collective, if a single CPU core is delayed by 1\,ms due to OS scheduling, all other GPUs busy-wait for that duration, amplifying a small per-core delay into a cluster-wide stall.

This \textit{straggler} phenomenon generalizes beyond collectives: any workload stage that involves CPU-side coordination, such as \texttt{DataLoader} workers, tokenization threads, or request-handling processes in serving engines, can trigger similar stalls when CPU cores are oversubscribed. Thus, provisioning CPU cores beyond the minimum process count is necessary to avoid these oversubscription-induced slowdowns. Analytical performance models have independently identified collective communication synchronization as a fundamental bottleneck in distributed LLM inference~\cite{davies2025liminal}; our measurements provide empirical confirmation of this on real hardware under CPU-constrained conditions.

\subsection{\REV{Kernel-Level Analysis of CPU-Induced Slowdown}}
\label{sec:kernel-profile}

\begin{figure}[t]
    \centering
    \includegraphics[width=1.0\linewidth]{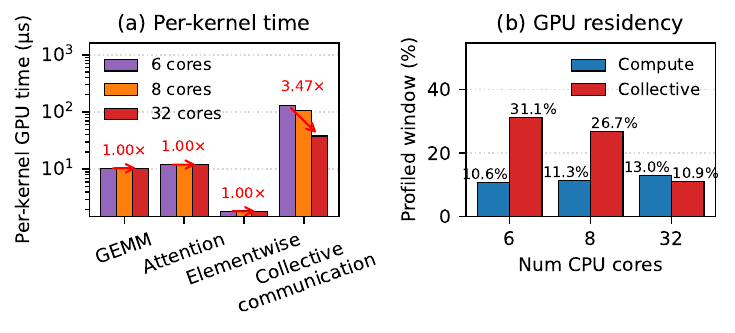}
    \vspace{-0.35in}
    \caption{\REV{Kernel-level profile under probe load (Llama 3.1 8B, TP=4, 16~RPS). (a)~GPU time per kernel. (b)~Fraction of kernel execution time relative to the profiled window of 300 engine iterations.}}
    \label{fig:kernel-profile}
    \vspace{-0.1in}
\end{figure}

To confirm that the slowdown does not originate from GPU execution, we profile individual GPU kernels from the ``probe under load'' experiment, averaged across ranks. \Cref{fig:kernel-profile}(a) shows that per-kernel execution time of compute kernels is unaffected by \#CPU: GEMM, attention, and elementwise kernels differ by less than 1\% between 6 and 32 cores. 
In contrast, the mean duration of communication collectives inflates by 3.47$\times$, and the inflation is asymmetric across ranks. 
For instance, at 6 cores, the p99 latency for the collective is 1.3\,s on rank~0 but 5.1--6.2\,s on the remaining ranks, which wait at the barrier for the delayed peer.
These results confirm that slower GPU compute execution is not the cause of the observed end-to-end slowdown~\cite{chung2025swift}.
 
\Cref{fig:kernel-profile}(b) decomposes GPU residency over the profiled window of 300 engine iterations. 
As CPU cores increase from 6 to 32, the time collectives spend spinning at synchronization barriers shrinks, and compute kernels take up a larger share of time within the window.
Because a spin-waiting collective still counts as "busy" in coarse utilization counters such as \texttt{nvidia-smi}, this decomposition reveals what such counters conceal: at 6 cores, most of the reported GPU activity is barrier spinning rather than useful work.

\begin{figure}[h]
    \centering
    \includegraphics[width=1.0\linewidth]{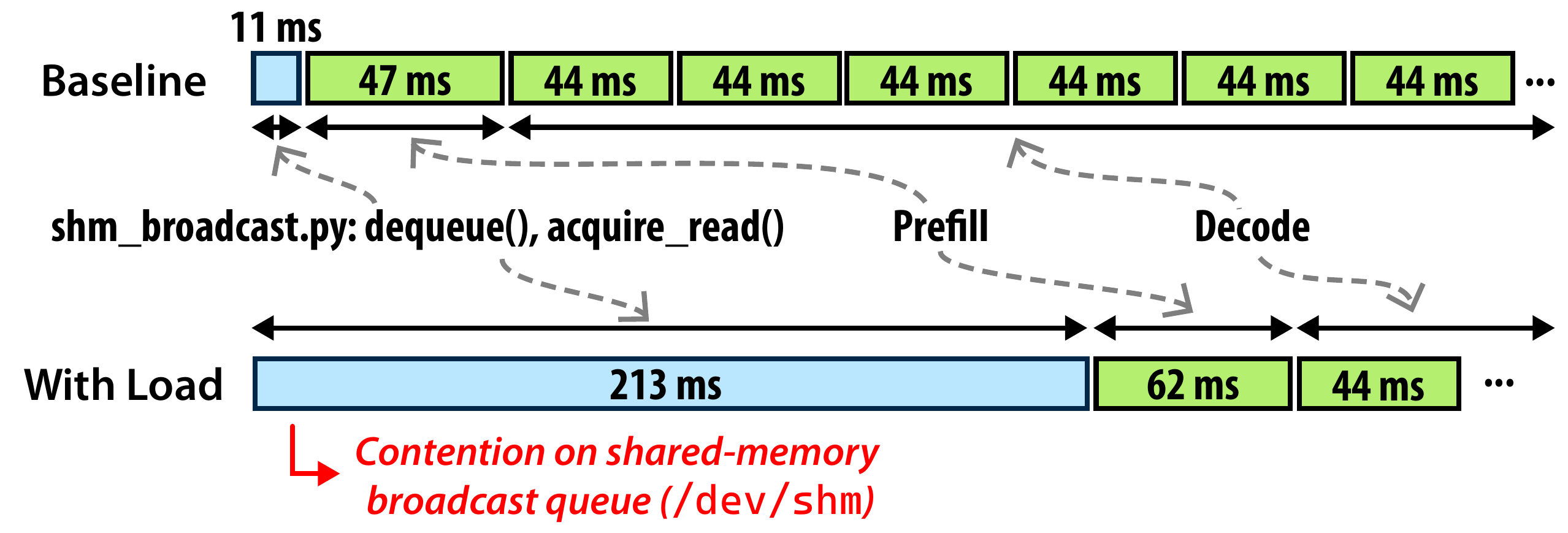}
    \vspace{-0.20in}
    \caption{Contended \texttt{dequeue()} on the shared-memory broadcast queue delays scheduling metadata to GPU workers (timing examples from Llama 3.1 8B).\label{fig:shmem}}
    \vspace{-0.1in}
\end{figure}

\subsection{Shared Memory Broadcast Contention}
\label{sec:shared}
Modern LLM serving frameworks such as vLLM~\cite{kwon2023efficient} and SGLang~\cite{zheng2024sglang} share a multi-process architecture in which the scheduler broadcasts work to all workers. 
We show that this broadcast introduces a second straggler effect, this time on the CPU side. The broadcast follows a 1-writer-$N$-reader pattern, where $N$ equals the number of GPUs.
The writer (engine core) must verify that all $N$ readers have consumed the previous message before writing new data, and each reader must confirm the writer has produced a new entry before consuming it. 
Although the usual implementation is lock-free (\eg vLLM), both the writer and the readers busy-wait without yielding to sleep. 
Under CPU scarcity, the writer's busy-wait competes with reader processes for CPU time, delaying readers' flag updates and forcing the writer to spin longer, creating a cascading delay that amplifies as the number of GPU workers increases.
\Cref{fig:shmem} illustrates how this contention can substantially delay model execution. 
In vLLM's V1 architecture, POSIX shared memory implements a broadcast queue that distributes scheduling decisions from EngineCore to each GPU worker process.
 
\begin{figure}[t]
    \centering
    \includegraphics[width=1.0\linewidth]{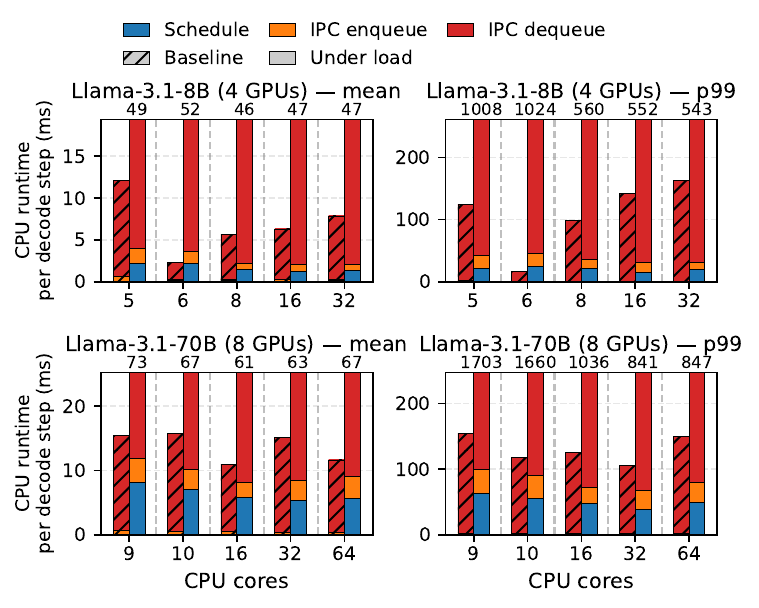}
    \vspace{-0.35in}
    \caption{CPU-side per-step runtime under baseline and load, decomposed into scheduling, IPC enqueue, and worker-side IPC dequeue. Rows show Llama 3.1 8B (4$\times$H100, TP=4) and Llama 3.1 70B (8$\times$H100, TP=8); columns show mean and p99 latency across CPU-core allocations.\label{fig:latency-breakdown}}
    \vspace{-0.1in}
\end{figure}

\noindent \textbf{Results.} \Cref{fig:latency-breakdown} decomposes the CPU-side work in a single vLLM V1 decode step under the same serving load used in \Cref{sec:inference}. 
We measure the latency of three host-side components: the engine scheduler, the engine-to-worker shared-memory enqueue, and the worker-side shared-memory dequeue.
The dequeue measurement includes time spent waiting for the next scheduling message. 
To avoid counting idle inter-arrival gaps as inter-process communication (IPC) overhead, we filter baseline dequeue waits above 200\,ms. 
The plotted time is therefore CPU-side control-plane time only; it excludes the GPU model forward pass.

The mean CPU-side per-step runtime grows from roughly \REV{2--16}\,ms at baseline to 46--73\,ms under load, depending on model size and CPU allocation. 
This increase persists even at larger CPU allocations, showing that CPU-side scheduling and IPC overheads remain significant under sustained load.
The tail is more severe: the component-wise p99 grows to 0.54--1.70\,s under load, with worker-side \texttt{dequeue()} dominating the CPU-side critical path (91--96\% of the p99 stack across all core allocations). 
This is a tail-latency effect rather than a shift in every iteration; typical steps remain much closer to baseline, but occasional delayed dequeues are enough to stall GPU workers and inflate request-level TTFT~\cite{dean2013tail}. 
Unlike the mean, the tail responds to CPU provisioning: p99 falls by roughly 2$\times$ from (\#GPUs + 1) cores to the most CPU-abundant allocation.
However, additional CPU cores only mitigate these tail stalls; they do not remove the underlying synchronization dependency in the broadcast path.

\REV{\noindent \textbf{Generality across serving frameworks.} We inspected the corresponding control paths in SGLang v0.5.16 and TensorRT-LLM v1.3.0, and both replicate the same one-writer, $N$-reader broadcast once per engine iteration: SGLang's rank-0 scheduler distributes batch metadata over a gloo CPU process group (\texttt{broadcast\_pyobj()}) in a loop that neither sleeps nor yields, and TensorRT-LLM's rank~0 issues an MPI broadcast over the tensor-parallel communicator every iteration. Only the transport differs; in all three systems, a delayed host process cannot consume the broadcast, and the back-pressure stalls subsequent iterations. Therefore, we claim that the bottleneck is a structural consequence of driving $N$ GPU workers from a replicated CPU control loop, independent of the engine or the actual implementation.}

\noindent \textbf{Takeaways.} While additional CPU cores mitigate oversubscription-induced slowdowns, broadcast contention is a structural property of the 1-writer-$N$-reader pattern that adding cores alone cannot fully resolve. Even with adequate CPU cores, the writer must wait for the slowest reader before proceeding. This wait also grows with the number of GPUs allocated to the same LLM engine. The issue compounds in multi-tenant environments where tokenization and IPC polling compete for the same CPU cores.

\noindent \textbf{Potential solutions.} Mitigating these bottlenecks may require redesigned IPC mechanisms, such as asynchronous scheduling pipelines that overlap IPC with GPU execution. Examples include persistent GPU kernels that poll a device-side queue to eliminate per-step launch overhead~\cite{liu2024deepseek, gupta2012study}, and direct GPU-to-GPU signaling that bypasses the CPU control plane. \REV{Each of these comes at a cost: a persistent-kernel design, for example, requires restructuring the serving engine around a finer-grained CPU--GPU dependency chain, since the host must continuously feed a device-side queue instead of issuing discrete per-step launches. We therefore leave the design of practical mechanisms to future work.}

\section{Discussion}
\label{sec:discussion}

\subsection{Case Study: CPU Underprovisioning in HPC (High-Performance Computing) Clusters}
\label{sec:slurm}

\begin{figure}[t]
    \centering
    \includegraphics[width=1.0\linewidth]{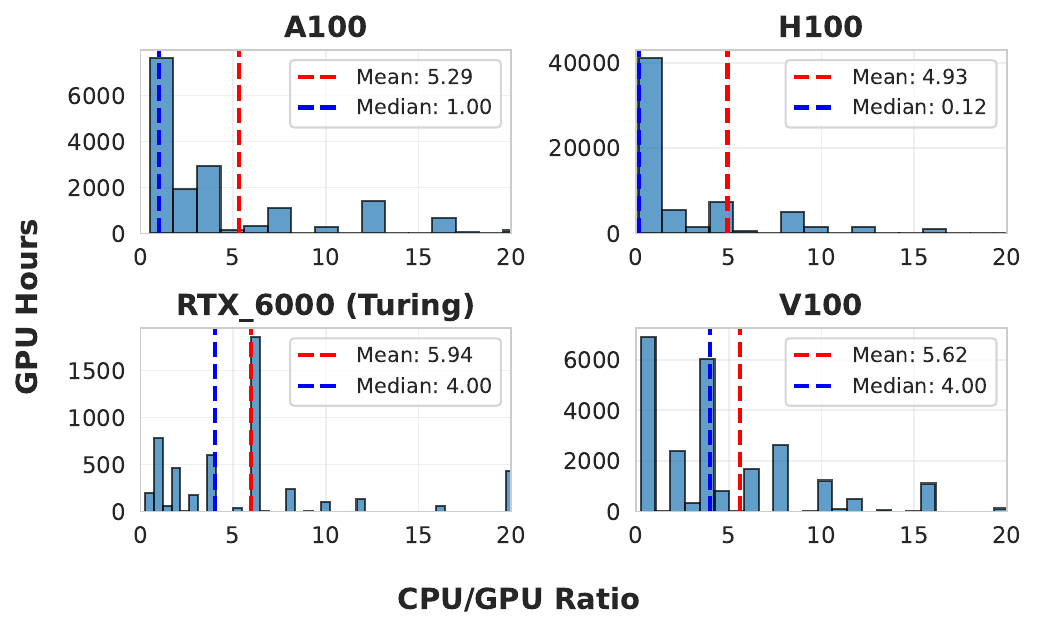}
    \vspace{-0.25in}
    \caption{GPU-hour-weighted CPU-to-GPU ratios on the instructional cluster (users manually set counts).\label{fig:pace-ice}}
    \vspace{-0.1in}
\end{figure}

As a concrete deployment case study, we analyze real-world resource allocation patterns from production HPC clusters. 
Prior cluster trace studies have characterized GPU job workloads at scale~\cite{hu2024characterization, weng2022mlaas, jeon2019analysis}, but have not specifically examined CPU-to-GPU allocation ratios. 
We find that CPUs are indeed frequently under-requested, suggesting that imbalances between CPUs and GPUs are a practical performance problem. 

We collected job scheduler logs from two institutional clusters operating within a university computing environment during calendar year 2024, totaling 4.65 million \texttt{salloc} records. The first cluster primarily serves research workloads, while the second supports instructional and educational use. Together, the clusters comprise approximately 1,400 compute nodes with around 35,000 CPU cores and several hundred GPUs, including recent high-end models such as NVIDIA H200 and RTX Pro 6000 Blackwell GPUs. \REV{The clusters serve heterogeneous CPU--GPU workloads, including ML training and inference as well as fluid-dynamics and computational-chemistry simulations, most of which run as multi-GPU jobs that also require multiple CPU workers per job.}

\begin{figure}[t]
    \centering
    \includegraphics[width=1.0\linewidth]{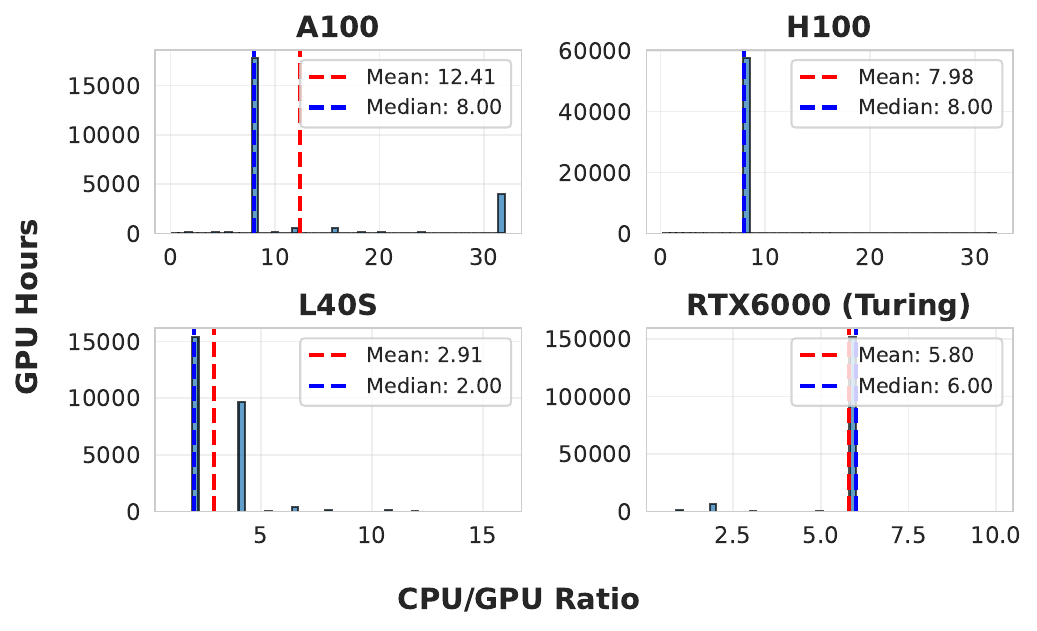}
    \vspace{-0.25in}
    \caption{GPU-hour-weighted CPU-to-GPU ratios on the research cluster (proportional unless overridden).\label{fig:pace-phoenix}}
    \vspace{-0.1in}
\end{figure}

\noindent \textbf{Resource Allocation Policies.} On the \textit{research cluster}, the scheduler enforces a fixed CPU-to-GPU allocation policy: each GPU receives roughly $1/N$ of the node's total CPU cores, where $N$ is the number of GPUs per node. 
This configuration preserves CPU/GPU affinity and prevents users from monopolizing CPU cores. 
In contrast, the \textit{instructional cluster} does not enforce a strict CPU-to-GPU ratio, reflecting its instructional goal: students are allowed to tune the allocation strategies. 
On this cluster, the default Slurm scheduling parameter \texttt{--cpus-per-task=1} allocates only a single CPU core per task, severely degrading the performance of multi-GPU workloads.

\noindent \textbf{Results.} \Cref{fig:pace-ice} summarizes the GPU-hour-weighted distribution of CPU-to-GPU allocation ratios in the \textit{instructional cluster}. We observe that many GPU-hours are concentrated at low CPU-to-GPU ratios, indicating that users often allocate far fewer CPU cores than GPUs. 
These low CPU allocations may partly reflect users forgetting or being unaware that additional CPU cores must be explicitly requested.
Notably, H100 nodes, accounting for 34.3k GPU hours out of 50.9k total, include cases in which users request as few as 1 CPU core across 4 or 8 GPUs. These findings suggest that many users remain unaware that inadequate CPU allocation can introduce severe bottlenecks in multi-GPU workloads.

\Cref{fig:pace-phoenix} shows the corresponding GPU-hour-weighted distribution for the \textit{research cluster}. Low CPU-to-GPU ratios are less pronounced because the scheduler enforces proportional CPU allocation. Nevertheless, the distribution still includes many allocations with fewer than 8 CPU cores per GPU. As our experiments in \Cref{sec:inference} show, ratios below 4--8 cores per GPU risk measurable slowdown for LLM inference workloads, leaving these jobs vulnerable to CPU-side performance bottlenecks.

\noindent \textbf{Takeaways.} While CPU scarcity is rare for large jobs that occupy an entire multi-GPU node (\eg a full DGX box), it is common in shared-node workloads where users run multi-GPU jobs with limited CPU allocations. The problem becomes more pronounced when the server operator fails to force the system to allocate sufficient CPU resources per requested GPU. These observations corroborate the CPU-induced bottlenecks identified in \Cref{sec:inference,sec:understanding}.

\subsection{CPU Underprovisioning in Cloud Compute Platforms}
\label{sec:cloud}

Public cloud providers such as AWS offer GPU instances with pricing that highlights a large cost disparity between CPU and GPU resources~\cite{aws}. GPU instances range from \$3.06/hour for a single V100 (p3.2xlarge) to \$55.04/hour for an 8$\times$H100 configuration (p5.48xlarge)~\cite{aws}. In contrast, CPU cores cost approximately \$0.03--\$0.06/hour per vCPU, making GPU compute roughly 100--200$\times$ more expensive than CPU cores depending on generation. Cloud instance configurations from major vendors commonly provide as few as 3--6 vCPUs per GPU~\cite{aws}, and the cluster-log case study above shows that users can allocate even fewer in shared environments.

Since the marginal cost of additional CPU cores is small relative to GPU instance pricing, \eg adding 16 vCPUs at \$0.05/hour each (\$0.80/hour) to a p5.48xlarge instance at \$55.04/hour represents roughly a 1.5\% cost increase, our analysis shows that adding CPU resources is remarkably cost-effective. For CPU-bound workloads, performance scales nearly linearly with additional cores at minimal cost, reducing the total GPU hours required and delivering better overall throughput per dollar than simply provisioning more GPUs. \REV{This cost argument extends to ARM-based hosts such as NVIDIA Grace. Grace Hopper systems remain host-bound across larger batch-size regimes than \texttt{x86} hosts~\cite{vellaisamy2025characterizing, fusco2024understanding}, so the CPU-side bottlenecks we characterize are likely to persist on these platforms as well. Meanwhile, ARM host cores carry a lower per-core cost than \texttt{x86} server cores, making CPU overprovisioning even more cost-effective relative to GPU pricing; vendors already ship generous ratios by default, such as 72 host cores per GPU on GH200~\cite{gracehopper}.}

We note that conventional host-side mitigation strategies do not fully address the CPU bottlenecks identified in this work.
Admission control and rate limiting can prevent CPU overload. 
However, they do so by reducing the accepted query rate, leaving GPU capacity underutilized and undermining the economics of multi-GPU deployment. 
OS-level resource isolation via \texttt{cgroup}, \texttt{cpuset}, and CPU pinning tools such as \texttt{taskset} and \texttt{isolcpus}~\cite{eunomia2025os} can improve scheduling determinism by dedicating cores to latency-sensitive processes, but cannot compensate when the total number of allocated cores is fundamentally insufficient for the workload's concurrency demands.
Accordingly, our evaluation focuses on aggregate CPU capacity rather than process-affinity policies.
While large cloud instances (\eg AWS p5.48xlarge) provision up to 24 vCPUs per GPU by default, multi-tenant sharing and cost-driven optimization frequently reduce effective per-GPU CPU allocations well below this level.

More fundamentally, solutions to CPU bottlenecks fall into two categories: \textit{host-driven} approaches that strengthen the CPU (adding cores, pinning, priority scheduling) and \textit{device-initiated} approaches that remove the CPU from the critical path entirely, such as GPU-initiated networking~\cite{nvshmem, si2025collective, hamidouche2025gpu, hwang2023ark}, storage access~\cite{gpudirect-storage, liu2025contention, qureshi2023gpu}, and persistent GPU kernels~\cite{liu2024deepseek}. Our results quantify the benefits of adding CPU cores; the latter represents a promising but more invasive architectural direction. Notably, modern GPUs already include on-die control processors, such as NVIDIA's RISC-V-based GPU System Processor (GSP)~\cite{nvidia-gsp}, which potentially can take over the orchestration duties currently handled by the host.

\subsection{Emerging Trends That Intensify the CPU Bottleneck}

Several emerging trends in LLM deployment suggest that CPU-side bottlenecks will become more pronounced rather than less. First, \textit{agentic AI} workloads perform dynamic reasoning with frequent tool invocations, each requiring CPU-side processing that creates substantial GPU idle time~\cite{kim2025cost, raj2025cpu, zheng2026agentcgroup, abhyankar2024infercept}. \REV{Moreover, \textit{long-context inference}, combined with \textit{frequent prefills} produced by multi-turn agentic tool invocations, imposes substantial load on the tokenizer, potentially consuming several seconds of CPU time for each request. As most newly introduced LLMs now natively support context lengths over 1M tokens, the tokenizer's CPU bottleneck will become even more significant~\cite{blakeman2025nvidia, team2026qwen3, kondrashov2026rethinking}. }

\textit{Multimodal inputs} such as images and video require CPU-intensive preprocessing (resizing, normalization, frame extraction) before GPU inference can begin, adding further pressure to the host. Finally, as GPU compute scales faster than CPU performance, deploying even moderately sized models across multiple GPUs reduces per-GPU compute time, increasing the relative contribution of CPU coordination overhead. Together, these trends indicate that the CPU bottlenecks characterized in this work are not artifacts of current-generation systems but structural challenges that will require architectural attention.

\subsection{Limitations}
Our study is bounded by the hardware and frameworks we could access. All experiments use NVIDIA GPUs with Intel CPUs and SMT disabled; production cloud instances typically expose vCPUs, so our physical-core measurements form a conservative baseline. 
Our findings may differ on CPU-GPU coupled architectures such as Grace Hopper~\cite{gracehopper}, where ARM cores show different kernel launch characteristics~\cite{vellaisamy2025characterizing}, or high-core-count AMD EPYC hosts.
Our quantitative evaluation centers on vLLM; while our structural analysis covers the control paths of SGLang~\cite{zheng2024sglang} and TensorRT-LLM~\cite{tensorrt-llm}, full quantitative validation across frameworks remains future work.

\section{Related Work}
\label{sec:related}

\subsection{CPU Overhead in Agentic AI}

Agentic AI workloads with dynamic reasoning and tool usage increasingly become CPU-bound~\cite{luo2025autellix, srivatsa2025preble}. Kim~\etalcite{kim2025cost} show that frequent tool invocations cause substantial GPU idle time, shifting workloads to CPU- or I/O-bound tasks. Raj~\etalcite{raj2025cpu} find that tool-processing components run predominantly on CPUs, accounting for up to 90.6\% of total latency.

\subsection{Host-Side LLM Inference Characterization}

Recent profiling studies quantify host-side overheads and scheduling challenges in agentic LLM inference~\cite{yang2026architectural, kondrashov2026rethinking}. Vellaisamy~\etalcite{vellaisamy2025characterizing} introduce SKIP profiler and the total kernel launch and queuing time (TKLQT) metric, showing that CPU-GPU coupled systems such as GH200 remain CPU-bound across larger batch-size regions where single-thread CPU performance and kernel launch overhead constrain low-batch latency. TaxBreak~\cite{vellaisamy2026taxbreak} decomposes host-visible orchestration overhead and shows that host-bound workloads, especially MoE inference, are sensitive to CPU single-thread performance; faster host CPUs reduce orchestration overhead by 10--29\% and improve end-to-end latency by up to 14\%. These studies diagnose single-node host overheads; our work is complementary, focusing on concurrent multi-GPU serving where tokenization, IPC broadcast, and barrier-based synchronization compound into TTFT slowdowns and timeouts. 

\subsection{GPU-Driven Control and Device-Initiated Communications}

GPU-driven execution and device-initiated communication aim to remove the CPU from the critical path. 
ARK~\cite{hwang2023ark} reduces CPU involvement by enabling GPU threads to directly control DMA operations.
NVSHMEM~\cite{nvshmem} enables device-initiated communication, with NCCLX~\cite{si2025collective} parallelizing remote direct memory access (RDMA) setup via lightweight GPU threads and DeepEP~\cite{liu2024deepseek} managing MoE expert parallelism~\cite{jiang2024mixtral}. Yet widely used frameworks such as vLLM and PyTorch still rely on CPU-driven coordination. LithOS~\cite{coppock2025lithos} redesigns the OS-GPU interface by atomizing long-running kernels for preemptive scheduling, reducing control-plane head-of-line blocking. Other work leverages spare CPU resources for distributed training~\cite{jiang2020unified}.

\subsection{LLM Serving Frameworks}

Modern LLM serving systems adopt multi-process architectures to isolate CPU-side tasks from GPU execution, including vLLM~\cite{kwon2023efficient} (\Cref{sec:setup}), SGLang~\cite{zheng2024sglang}, and TensorRT-LLM~\cite{tensorrt-llm}. DistServe~\cite{zhong2024distserve} and Splitwise~\cite{patel2024splitwise} disaggregate the prefill and decode phases onto separate GPU pools to optimize goodput, thereby concentrating CPU-intensive tokenization on dedicated prefill servers. While these frameworks optimize GPU utilization and scheduling, all rely on CPU-side processes for tokenization, request management, and inter-process coordination, leaving them vulnerable to the bottlenecks characterized in this work.

\subsection{Mitigating the CPU Overhead in LLMs}

Tokenization is a key CPU bottleneck in LLMs. NetTokenizer~\cite{mehta2025accelerating} offloads tokenization to programmable network data planes, BlockBPE~\cite{you2025blockbpe} runs BPE on GPUs (effective for high-batch but not latency-sensitive workloads), and LoPT~\cite{shao2025lopt} achieves lossless parallel tokenization via token merging. Recasens~\etalcite{recasens2025mind} report CPU overhead reaching up to 30\% of total execution time in large-batch LLM inference.

\subsection{Mitigating Kernel Launch Overhead}

Lustig~\etalcite{lustig2013reducing} propose early kernel launch to reduce CPU-GPU synchronization latency, and CPU pinning reduces scheduling jitter~\cite{eunomia2025os}. CUDA Graphs~\cite{cudagraph, ekelund2025boosting} reduce launch overhead for static sequences but cannot capture dynamic LLM decoding decisions; Grape~\cite{zheng2023grape} and KTransformers~\cite{chen2025ktransformers} extend them to dynamic DNNs and MoE models, respectively, but neither addresses tokenization contention. FlashAttention~\cite{dao2022flashattention} fuses kernels but does not address launch-path overhead.

Prior works have independently noted CPU overhead in tokenization~\cite{you2025blockbpe, shao2025lopt}, host orchestration and kernel launch paths~\cite{hwang2023ark, vellaisamy2025characterizing, vellaisamy2026taxbreak}, and data preprocessing pipelines~\cite{mohan2022looking}, while complementary GPU-side characterizations~\cite{wang2025systematic, recasens2025mind} focus on compute and memory bottlenecks within the accelerator. None of these works examines how these factors compound under multi-GPU LLM serving, where tokenization contention, IPC broadcast delays, and barrier-based synchronization interact to produce cascading slowdowns. Our work systematically characterizes the CPU-side effects that compound with these GPU-side phenomena and quantifies how CPU provisioning affects end-to-end performance.

\section{Conclusion}
\label{sec:conclusion}

Our study shows that CPU-side responsibilities can become critical bottlenecks in multi-GPU LLM inference, even in fully optimized serving stacks. CPU oversubscription, combined with barrier-based GPU synchronization, leaves GPUs idle, while shared-memory broadcast contention introduces structural delays proportional to the degree of tensor parallelism. Providing adequate CPU resources improves TTFT by 1.47--7.11$\times$ at little marginal cost relative to GPUs, and our analysis of 4.65 million cluster job records confirms widespread underprovisioning.

\section*{Acknowledgments}

We thank the anonymous reviewers for their insightful feedback. 
This work was supported in part by an NSF grant CCF-2316176.
Claude Opus 4.7 was used for grammar checking and assistance with experimental environment setup; the authors take full responsibility for all content and results.

\bibliographystyle{IEEEtranS}
\bibliography{reference}

@misc{dgx-b200,
    organization={NVIDIA},
    title={NVIDIA DGX B200},
    howpublished = {\url{https://www.nvidia.com/en-us/data-center/dgx-b200/}}
}

@misc{dgx-h100,
    organization={NVIDIA},
    title={NVIDIA DGX H100},
    howpublished = {\url{https://www.nvidia.com/en-gb/data-center/dgx-h100/}}
}

@misc{nccl,
    organization={NVIDIA},
    title={NVIDIA Collective Communications Library (NCCL)},
    howpublished={\url{https://docs.nvidia.com/deeplearning/nccl/user-guide/docs/usage/collectives.html}}
}

@article{li2020pytorch,
  title={Pytorch distributed: Experiences on accelerating data parallel training},
  author={Li, Shen and Zhao, Yanli and Varma, Rohan and Salpekar, Omkar and Noordhuis, Pieter and Li, Teng and Paszke, Adam and Smith, Jeff and Vaughan, Brian and Damania, Pritam and others},
  journal={arXiv preprint arXiv:2006.15704},
  year={2020}
}

@article{shoeybi2019megatron,
  title={Megatron-lm: Training multi-billion parameter language models using model parallelism},
  author={Shoeybi, Mohammad and Patwary, Mostofa and Puri, Raul and LeGresley, Patrick and Casper, Jared and Catanzaro, Bryan},
  journal={arXiv preprint arXiv:1909.08053},
  year={2019}
}

@misc{nvidia-gsp,
  title = {{NVIDIA} {GSP} Firmware},
  howpublished = {\url{https://download.nvidia.com/XFree86/Linux-x86_64/560.28.03/README/gsp.html}},
  year = {2024}
  }

@article{liu2024deepseek,
  title={Deepseek-v3 technical report},
  author={Liu, Aixin and Feng, Bei and Xue, Bing and Wang, Bingxuan and Wu, Bochao and Lu, Chengda and Zhao, Chenggang and Deng, Chengqi and Zhang, Chenyu and Ruan, Chong and others},
  journal={arXiv preprint arXiv:2412.19437},
  year={2024}
}

@article{hamidouche2025gpu,
  title={GPU-Initiated Networking for NCCL},
  author={Hamidouche, Khaled and Bachan, John and Markthub, Pak and Gootzen, Peter-Jan and Agostini, Elena and Jeaugey, Sylvain and Shafi, Aamir and Theodorakis, Georgios and Venkata,
Manjunath Gorentla},
  journal={arXiv preprint arXiv:2511.15076},
  year={2025}
}

@article{zheng2026agentcgroup,
  title={AgentCgroup: Understanding and Controlling OS Resources of AI Agents},
  author={Zheng, Yusheng and Fan, Jiakun and Fu, Quanzhi and Yang, Yiwei and Zhang, Wei and Quinn, Andi},
  journal={arXiv preprint arXiv:2602.09345},
  year={2026}
}

@inproceedings{kwon2023efficient,
  title={Efficient memory management for large language model serving with pagedattention},
  author={Kwon, Woosuk and Li, Zhuohan and Zhuang, Siyuan and Sheng, Ying and Zheng, Lianmin and Yu, Cody Hao and Gonzalez, Joseph and Zhang, Hao and Stoica, Ion},
  booktitle={Proceedings of the 29th symposium on operating systems principles},
  pages={611--626},
  year={2023}
}

@misc{wu2020visual,
      title={Visual Transformers: Token-based Image Representation and Processing for Computer Vision}, 
      author={Bichen Wu and Chenfeng Xu and Xiaoliang Dai and Alvin Wan and Peizhao Zhang and Zhicheng Yan and Masayoshi Tomizuka and Joseph Gonzalez and Kurt Keutzer and Peter Vajda},
      year={2020},
      eprint={2006.03677},
      archivePrefix={arXiv},
      primaryClass={cs.CV}
}

@misc{touvron2021training-deit,
      title={Training data-efficient image transformers \& distillation through attention}, 
      author={Hugo Touvron and Matthieu Cord and Matthijs Douze and Francisco Massa and Alexandre Sablayrolles and Hervé Jégou},
      year={2021},
      eprint={2012.12877},
      archivePrefix={arXiv},
      primaryClass={cs.CV}
}

@article{recasens2025mind,
  title={Mind the memory gap: Unveiling gpu bottlenecks in large-batch llm inference},
  author={Recasens, Pol G and Agullo, Ferran and Zhu, Yue and Wang, Chen and Lee, Eun Kyung and Tardieu, Olivier and Torres, Jordi and Berral, Josep Ll},
  journal={arXiv preprint arXiv:2503.08311},
  year={2025}
}

@inproceedings{jiang2020unified,
  title={A unified architecture for accelerating distributed $\{$DNN$\}$ training in heterogeneous $\{$GPU/CPU$\}$ clusters},
  author={Jiang, Yimin and Zhu, Yibo and Lan, Chang and Yi, Bairen and Cui, Yong and Guo, Chuanxiong},
  booktitle={14th USENIX Symposium on Operating Systems Design and Implementation (OSDI 20)},
  pages={463--479},
  year={2020}
}

@misc{scaling,
    title={Scaling Analysis},
    organization={Princeton Research Computing},
    howpublished={\url{https://researchcomputing.princeton.edu/support/knowledge-base/scaling-analysis}}
}

@misc{slurm,
    title={Slurm documentation},
    organization={Slurm},
    howpublished={\url{https://slurm.schedmd.com/documentation.html}}
}

@inproceedings{weng2022mlaas,
  title={$\{$MLaaS$\}$ in the wild: Workload analysis and scheduling in $\{$Large-Scale$\}$ heterogeneous $\{$GPU$\}$ clusters},
  author={Weng, Qizhen and Xiao, Wencong and Yu, Yinghao and Wang, Wei and Wang, Cheng and He, Jian and Li, Yong and Zhang, Liping and Lin, Wei and Ding, Yu},
  booktitle={19th USENIX Symposium on Networked Systems Design and Implementation (NSDI 22)},
  pages={945--960},
  year={2022}
}

@inproceedings{zhao2025insights,
  title={Insights into deepseek-v3: Scaling challenges and reflections on hardware for ai architectures},
  author={Zhao, Chenggang and Deng, Chengqi and Ruan, Chong and Dai, Damai and Gao, Huazuo and Li, Jiashi and Zhang, Liyue and Huang, Panpan and Zhou, Shangyan and Ma, Shirong and others},
  booktitle={Proceedings of the 52nd Annual International Symposium on Computer Architecture},
  pages={1731--1745},
  year={2025}
}

@misc{huggingface-tokenizers,
    title={HuggingFace Tokenizers},
    organization={HuggingFace},
    howpublished={\url{https://huggingface.co/docs/tokenizers/index}}}

@misc{DALI,
  organization = {NVIDIA},
  title = {NVIDIA DALI: A GPU-accelerated data loading and pre-processing library},
  howpublished = {\url{https://github.com/NVIDIA/DALI}},
  year = {2017}
}

@article{krizhevsky2012imagenet,
  title={Imagenet classification with deep convolutional neural networks},
  author={Krizhevsky, Alex and Sutskever, Ilya and Hinton, Geoffrey E},
  journal={Advances in neural information processing systems},
  volume={25},
  year={2012}
}

@article{raj2025cpu,
  title={A CPU-Centric Perspective on Agentic AI},
  author={Raj, Ritik and Wang, Hong and Krishna, Tushar},
  journal={arXiv preprint arXiv:2511.00739},
  year={2025}
}

@inproceedings{hu2024characterization,
  title={Characterization of large language model development in the datacenter},
  author={Hu, Qinghao and Ye, Zhisheng and Wang, Zerui and Wang, Guoteng and Zhang, Meng and Chen, Qiaoling and Sun, Peng and Lin, Dahua and Wang, Xiaolin and Luo, Yingwei and others},
  booktitle={21st USENIX Symposium on Networked Systems Design and Implementation (NSDI 24)},
  pages={709--729},
  year={2024}
}

@article{davies2025liminal,
  title={LIMINAL: Exploring The Frontiers of LLM Decode Performance},
  author={Davies, Michael and Crago, Neal and Sankaralingam, Karthikeyan and Kozyrakis, Christos},
  journal={arXiv preprint arXiv:2507.14397},
  year={2025}
}

@article{wang2025systematic,
  title={A Systematic Characterization of LLM Inference on GPUs},
  author={Wang, Haonan and Xiao, Xuxin and Yan, Mingyu and Zhu, Zhuoyuan and Han, Dengke and Wang, Duo and Li, Wenming and Ye, Xiaochun and Hu, Cunchen and Chen, Hongyang and others},
  journal={arXiv preprint arXiv:2512.01644},
  year={2025}
}

@article{kim2025cost,
  title={The Cost of Dynamic Reasoning: Demystifying AI Agents and Test-Time Scaling from an AI Infrastructure Perspective},
  author={Kim, Jiin and Shin, Byeongjun and Chung, Jinha and Rhu, Minsoo},
  journal={arXiv preprint arXiv:2506.04301},
  year={2025}
}

@inproceedings{rasley2020deepspeed,
  title={Deepspeed: System optimizations enable training deep learning models with over 100 billion parameters},
  author={Rasley, Jeff and Rajbhandari, Samyam and Ruwase, Olatunji and He, Yuxiong},
  booktitle={Proceedings of the 26th ACM SIGKDD international conference on knowledge discovery \& data mining},
  pages={3505--3506},
  year={2020}
}

@misc{accelerate,
  title =        {Accelerate: Training and inference at scale made simple, efficient and adaptable.},
  author =       {Sylvain Gugger and Lysandre Debut and Thomas Wolf and Philipp Schmid and Zachary Mueller and Sourab Mangrulkar and Marc Sun and Benjamin Bossan},
  howpublished = {\url{https://github.com/huggingface/accelerate}},
  year =         {2022}
}

@article{si2025collective,
  title={Collective Communication for 100k+ GPUs},
  author={Si, Min and Balaji, Pavan and Chen, Yongzhou and Chu, Ching-Hsiang and Gangidi, Adi and Hasan, Saif and Iyengar, Subodh and Johnson, Dan and Liu, Bingzhe and Ren, Jingliang and others},
  journal={arXiv preprint arXiv:2510.20171},
  year={2025}
}

@inproceedings{hwang2023ark,
  title={$\{$ARK$\}$:$\{$GPU-driven$\}$ code execution for distributed deep learning},
  author={Hwang, Changho and Park, KyoungSoo and Shu, Ran and Qu, Xinyuan and Cheng, Peng and Xiong, Yongqiang},
  booktitle={20th USENIX Symposium on Networked Systems Design and Implementation (NSDI 23)},
  pages={87--101},
  year={2023}
}

@misc{nvshmem,
    organization={NVIDIA},
    title={NVSHMEM},
    howpublished = {\url{https://developer.nvidia.com/nvshmem}},
    year = {2024}
}

@inproceedings{mohan2022looking,
  title={Looking beyond $\{$GPUs$\}$ for $\{$DNN$\}$ scheduling on $\{$Multi-Tenant$\}$ clusters},
  author={Mohan, Jayashree and Phanishayee, Amar and Kulkarni, Janardhan and Chidambaram, Vijay},
  booktitle={16th USENIX Symposium on Operating Systems Design and Implementation (OSDI 22)},
  pages={579--596},
  year={2022}
}

@misc{aws,
    organization={AWS},
    title={Amazon EC2 instance types},
    howpublished={\url{https://aws.amazon.com/ec2/instance-types/}}
}

@article{mehta2025accelerating,
  title={Accelerating LLM Inference with Smart NIC Tokenization and Caching},
  author={Mehta, Sudarshan},
  year={2025},
  publisher={Santa Clara: Santa Clara University, 2025}
}

@misc{shao2025lopt,
      title={LoPT: Lossless Parallel Tokenization Acceleration for Long Context Inference of Large Language Model}, 
      author={Wei Shao and Lingchao Zheng and Pengyu Wang and Peizhen Zheng and Jun Li and Yuwei Fan},
      year={2025},
      eprint={2511.04952},
      archivePrefix={arXiv},
      primaryClass={cs.CL},
      url={https://arxiv.org/abs/2511.04952}, 
}

@article{you2025blockbpe,
  title={BlockBPE: Parallel BPE Tokenization},
  author={You, Amos},
  journal={arXiv preprint arXiv:2507.11941},
  year={2025}
}

@misc{cudagraph,
    title={Getting Started with CUDA Graphs},
    organization={NVIDIA},
    howpublished={\url{https://developer.nvidia.com/blog/cuda-graphs/}}
}

@article{hu2025demystifying,
  title={Demystifying NCCL: An in-depth analysis of GPU communication protocols and algorithms},
  author={Hu, Zhiyi and Shen, Siyuan and Bonato, Tommaso and Jeaugey, Sylvain and Alexander, Cedell and Spada, Eric and Dinan, James and Hammond, Jeff and Hoefler, Torsten},
  journal={arXiv preprint arXiv:2507.04786},
  year={2025}
}

@inproceedings{vellaisamy2025characterizing,
  title={Characterizing and optimizing llm inference workloads on cpu-gpu coupled architectures},
  author={Vellaisamy, Prabhu and Labonte, Thomas and Chakraborty, Sourav and Turner, Matt and Sury, Samantika and Shen, John Paul},
  booktitle={2025 IEEE International Symposium on Performance Analysis of Systems and Software (ISPASS)},
  pages={49--61},
  year={2025},
  organization={IEEE}
}

@inproceedings{sennrich2016neural,
  title={Neural machine translation of rare words with subword units},
  author={Sennrich, Rico and Haddow, Barry and Birch, Alexandra},
  booktitle={Proceedings of the 54th annual meeting of the association for computational linguistics (volume 1: long papers)},
  pages={1715--1725},
  year={2016}
}

@article{kudo2018sentencepiece,
  title={SentencePiece: A simple and language independent subword tokenizer and detokenizer for neural text processing},
  author={Kudo, Taku and Richardson, John},
  journal={arXiv preprint arXiv:1808.06226},
  year={2018}
}

@misc{pytorch-profiler,
    title={torch.profiler -- PyTorch 2.9 documentation},
    organization={Pytorch},
    howpublished={\url{https://docs.pytorch.org/docs/stable/profiler.html}}
}

@article{durvasula2024acs,
  title={ACS: Concurrent Kernel Execution on Irregular, Input-Dependent Computational Graphs},
  author={Durvasula, Sankeerth and Zhao, Adrian and Kiguru, Raymond and Guan, Yushi and Chen, Zhonghan and Vijaykumar, Nandita},
  journal={arXiv preprint arXiv:2401.12377},
  year={2024}
}

@article{wang2024beyond,
  title={Beyond the limits: A survey of techniques to extend the context length in large language models},
  author={Wang, Xindi and Salmani, Mahsa and Omidi, Parsa and Ren, Xiangyu and Rezagholizadeh, Mehdi and Eshaghi, Armaghan},
  journal={arXiv preprint arXiv:2402.02244},
  year={2024}
}

@inproceedings{coppock2025lithos,
  title={LithOS: An operating system for efficient machine learning on GPUs},
  author={Coppock, Patrick H and Zhang, Brian and Solomon, Eliot H and Kypriotis, Vasilis and Yang, Leon and Sharma, Bikash and Schatzberg, Dan and Mowry, Todd C and Skarlatos, Dimitrios},
  booktitle={Proceedings of the ACM SIGOPS 31st Symposium on Operating Systems Principles},
  pages={1--17},
  year={2025}
}

@misc{gracehopper,
    title={NVIDIA GH200 Grace Hopper Superchip},
    organization={NVIDIA},
    howpublished={\url{https://www.nvidia.com/en-us/data-center/grace-hopper-superchip/}}
}

@article{dao2022flashattention,
  title={Flashattention: Fast and memory-efficient exact attention with io-awareness},
  author={Dao, Tri and Fu, Dan and Ermon, Stefano and Rudra, Atri and R{\'e}, Christopher},
  journal={Advances in neural information processing systems},
  volume={35},
  pages={16344--16359},
  year={2022}
}

@inproceedings{lustig2013reducing,
  title={Reducing GPU offload latency via fine-grained CPU-GPU synchronization},
  author={Lustig, Daniel and Martonosi, Margaret},
  booktitle={2013 IEEE 19th International Symposium on High Performance Computer Architecture (HPCA)},
  pages={354--365},
  year={2013},
  organization={IEEE}
}

@inproceedings{zheng2023grape,
  title={Grape: Practical and efficient graphed execution for dynamic deep neural networks on gpus},
  author={Zheng, Bojian and Yu, Cody Hao and Wang, Jie and Ding, Yaoyao and Liu, Yizhi and Wang, Yida and Pekhimenko, Gennady},
  booktitle={Proceedings of the 56th Annual IEEE/ACM International Symposium on Microarchitecture},
  pages={1364--1380},
  year={2023}
}

@inproceedings{chen2025ktransformers,
  title={Ktransformers: Unleashing the full potential of cpu/gpu hybrid inference for moe models},
  author={Chen, Hongtao and Xie, Weiyu and Zhang, Boxin and Tang, Jingqi and Wang, Jiahao and Dong, Jianwei and Chen, Shaoyuan and Yuan, Ziwei and Lin, Chen and Qiu, Chengyu and others},
  booktitle={Proceedings of the ACM SIGOPS 31st Symposium on Operating Systems Principles},
  pages={1014--1029},
  year={2025}
}

@misc{tensorrt-llm,
    title={TensorRT-LLM},
    organization={NVIDIA},
    howpublished={\url{https://github.com/NVIDIA/TensorRT-LLM}},
    year={2023}
}

@article{zheng2024sglang,
  title={Sglang: Efficient execution of structured language model programs},
  author={Zheng, Lianmin and Yin, Liangsheng and Xie, Zhiqiang and Sun, Chuyue and Huang, Jeff and Yu, Cody H and Cao, Shiyi and Kozyrakis, Christos and Stoica, Ion and Gonzalez, Joseph E and others},
  journal={Advances in neural information processing systems},
  volume={37},
  pages={62557--62583},
  year={2024}
}

@inproceedings{zhong2024distserve,
  title={$\{$DistServe$\}$: Disaggregating prefill and decoding for goodput-optimized large language model serving},
  author={Zhong, Yinmin and Liu, Shengyu and Chen, Junda and Hu, Jianbo and Zhu, Yibo and Liu, Xuanzhe and Jin, Xin and Zhang, Hao},
  booktitle={18th USENIX Symposium on Operating Systems Design and Implementation (OSDI 24)},
  pages={193--210},
  year={2024}
}

@inproceedings{patel2024splitwise,
  title={Splitwise: Efficient generative llm inference using phase splitting},
  author={Patel, Pratyush and Choukse, Esha and Zhang, Chaojie and Shah, Aashaka and Goiri, {\'I}{\~n}igo and Maleki, Saeed and Bianchini, Ricardo},
  booktitle={2024 ACM/IEEE 51st Annual International Symposium on Computer Architecture (ISCA)},
  pages={118--132},
  year={2024},
  organization={IEEE}
}

@article{jiang2024mixtral,
  title={Mixtral of experts},
  author={Jiang, Albert Q and Sablayrolles, Alexandre and Roux, Antoine and Mensch, Arthur and Savary, Blanche and Bamford, Chris and Chaplot, Devendra Singh and Casas, Diego de las and Hanna, Emma Bou and Bressand, Florian and others},
  journal={arXiv preprint arXiv:2401.04088},
  year={2024}
}

@article{wong2025gpus,
  title={GPUs, CPUs, and... NICs: Rethinking the Network's Role in Serving Complex AI Pipelines},
  author={Wong, Mike and Butler, Ulysses and Farkash, Emma and Tammana, Praveen and Sivaraman, Anirudh and Netravali, Ravi},
  journal={arXiv preprint arXiv:2502.15712},
  year={2025}
}

@inproceedings{ekelund2025boosting,
  title={Boosting performance of iterative applications on gpus: Kernel batching with cuda graphs},
  author={Ekelund, Jonah and Markidis, Stefano and Peng, Ivy},
  booktitle={2025 33rd Euromicro International Conference on Parallel, Distributed, and Network-Based Processing (PDP)},
  pages={70--77},
  year={2025},
  organization={IEEE}
}

@misc{eunomia2025os,
      title = {OS-Level Challenges in {LLM} Inference and Optimizations},
      howpublished = {\url{https://eunomia.dev/blog/2025/02/18/os-level-challenges-in-llm-inference-and-optimizations/}},
      organization={Eunomia},
      year = {2025}
  }

@article{kondrashov2026rethinking,
  title={Rethinking AI Cloud Infrastructure for Agentic Serving Systems with the Aries Experimentation Framework},
  author={Kondrashov, Leonid and Liu, Hongrui and Park, JooYoung and Zhou, Boxi and Liu, Zonghao and Lu, Chengzhi and Mancini, Riccardo and Choukse, Esha and Javaid, Haris and Sviridov, German and others},
  journal={arXiv preprint arXiv:2607.29069},
  year={2026}
}

@article{zhang2026toktier,
  title={TokTier: Exact Stateful Tokenization for Agentic LLM Serving},
  author={Zhang, Zhenyu and Cao, Zhichao},
  journal={arXiv preprint arXiv:2607.29678},
  year={2026}
}

@inproceedings{xiang2026servegen,
  title={$\{$ServeGen$\}$: Workload Characterization and Generation of Large Language Model Serving in Production},
  author={Xiang, Yuxing and Li, Xue and Qian, Kun and Zhang, Yan and Yu, Wenyuan and Zhai, Ennan and Jin, Xin and Zhou, Jingren},
  booktitle={23rd USENIX Symposium on Networked Systems Design and Implementation (NSDI 26)},
  pages={1845--1859},
  year={2026}
}

@article{wang2024burstgpt,
  title={BurstGPT: A Real-world Workload Dataset to Optimize LLM Serving Systems.(2024)},
  author={Wang, Yuxin and Chen, Yuhan and Li, Zeyu and Kang, Xueze and Tang, Zhenheng and He, Xin and Guo, Rui and Wang, Xin and Wang, Qiang and Zhou, Amelie Chi and others},
  journal={URL https://arxiv. org/abs/2401},
  volume={17644},
  year={2024}
}

@inproceedings{srivatsa2025preble,
  title={Preble: Efficient distributed prompt scheduling for llm serving},
  author={Srivatsa, Vikranth and He, Zijian and Abhyankar, Reyna and Li, Dongming and Zhang, Yiying},
  booktitle={International conference on learning representations},
  volume={2025},
  pages={37057--37082},
  year={2025}
}

@inproceedings{yu2025stateful,
  title={Stateful large language model serving with pensieve},
  author={Yu, Lingfan and Lin, Jinkun and Li, Jinyang},
  booktitle={Proceedings of the Twentieth European Conference on Computer Systems},
  pages={144--158},
  year={2025}
}

@article{li2025continuum,
  title={Continuum: Efficient and robust multi-turn llm agent scheduling with kv cache time-to-live},
  author={Li, Hanchen and He, Runyuan and Mang, Qiuyang and Zhang, Qizheng and Mao, Huanzhi and Chen, Xiaokun and Zhou, Hangrui and Cheung, Alvin and Gonzalez, Joseph and Stoica, Ion},
  journal={arXiv preprint arXiv:2511.02230},
  year={2025}
}

@article{chang2026llm,
  title={From LLM Inference to Agentic Workloads: Characterization and Implications for Serving Systems},
  author={Chang, Chaokun and Zhou, Yukun and Fu, Kaihua and An, Dakai and Feng, Tianyu and Lu, Hanfeng and Yao, Sheng and Guo, Pu and Yu, Yinghao and Shan, Yizhou and others},
  journal={arXiv preprint arXiv:2608.15127},
  year={2026}
}

@article{liu2026agentic,
  title={Agentic Coding in the Wild: Characterizing GitHub Copilot Traces at Production Scale},
  author={Liu, Banruo and Qiu, Haoran and Goiri, {\'I}{\~n}igo and Fonseca, Rodrigo and Bianchini, Ricardo and Choukse, Esha},
  journal={arXiv preprint arXiv:2608.00101},
  year={2026}
}

@inproceedings{vellaisamy2026taxbreak,
  title={TaxBreak: Unmasking the Hidden Costs of LLM Inference Through Overhead Decomposition},
  author={Vellaisamy, Prabhu and Tripathi, Shreesh and Natarajan, Vignesh and Thenarasu, Surya Santhan and Blanton, Shawn and Shen, John P},
  booktitle={2026 IEEE International Symposium on Performance Analysis of Systems and Software (ISPASS)},
  year={2026},
  organization={IEEE}
}

@misc{rising2026varra,
      title = {The Rising CPU:GPU Ratio in AI Infrastructure: Drivers, Trends, and Implications},
      author={Ram Varra and Sachin Ashtikar and Vipul Lal and Shesha Krishnapura},
      organization={Intel},
      year = {2026}
  }

@article{qin2024mooncake,
  title={Mooncake: A kvcache-centric disaggregated architecture for llm serving},
  author={Qin, Ruoyu and Li, Zheming and He, Weiran and Cui, Jialei and Tang, Heyi and Ren, Feng and Ma, Teng and Cai, Shangming and Zhang, Yineng and Zhang, Mingxing and others},
  journal={ACM Transactions on Storage},
  year={2024},
  publisher={ACM New York, NY}
}

@article{team2026qwen3,
  title={Qwen3. 5-omni technical report},
  author={Team, Qwen},
  journal={arXiv preprint arXiv:2604.15804},
  year={2026}
}

@misc{fastokens,
    title={fastokens},
    author={{Crusoe Cloud}},
    year={2026},
    howpublished={\url{https://github.com/crusoecloud/fastokens}}
}

@inproceedings{qureshi2023gpu,
  title={Gpu-initiated on-demand high-throughput storage access in the bam system architecture},
  author={Qureshi, Zaid and Mailthody, Vikram Sharma and Gelado, Isaac and Min, Seungwon and Masood, Amna and Park, Jeongmin and Xiong, Jinjun and Newburn, Chris J and Vainbrand, Dmitri and Chung, I-Hsin and others},
  booktitle={Proceedings of the 28th ACM International Conference on Architectural Support for Programming Languages and Operating Systems, Volume 2},
  pages={325--339},
  year={2023}
}

@article{yang2026architectural,
  title={Architectural Implications of Agentic AI Workflows},
  author={Yang, Jirong and Liu, Peizhe and Zhang, Chaojie and Stojkovic, Jovan},
  journal={arXiv preprint arXiv:2608.04458},
  year={2026}
}

@misc{gpudirect-storage,
    title={NVIDIA GPUDirect Storage},
    author={{NVIDIA}},
    year={2024},
    howpublished={\url{https://docs.nvidia.com/gpudirect-storage/}}
}

@article{luo2025autellix,
  title={Autellix: An efficient serving engine for llm agents as general programs},
  author={Luo, Michael and Shi, Xiaoxiang and Cai, Colin and Zhang, Tianjun and Wong, Justin and Wang, Yichuan and Wang, Chi and Huang, Yanping and Chen, Zhifeng and Gonzalez, Joseph E and others},
  journal={arXiv preprint arXiv:2502.13965},
  year={2025}
}

@article{abhyankar2024infercept,
  title={Infercept: Efficient intercept support for augmented large language model inference},
  author={Abhyankar, Reyna and He, Zijian and Srivatsa, Vikranth and Zhang, Hao and Zhang, Yiying},
  journal={arXiv preprint arXiv:2402.01869},
  year={2024}
}

@inproceedings{gujarati2020serving,
  title={Serving $\{$DNNs$\}$ like clockwork: Performance predictability from the bottom up},
  author={Gujarati, Arpan and Karimi, Reza and Alzayat, Safya and Hao, Wei and Kaufmann, Antoine and Vigfusson, Ymir and Mace, Jonathan},
  booktitle={14th USENIX Symposium on Operating Systems Design and Implementation (OSDI 20)},
  pages={443--462},
  year={2020}
}

@article{fusco2024understanding,
  title={Understanding data movement in tightly coupled heterogeneous systems: A case study with the grace hopper superchip},
  author={Fusco, Luigi and Khalilov, Mikhail and Chrapek, Marcin and Chukkapalli, Giridhar and Schulthess, Thomas and Hoefler, Torsten},
  journal={arXiv preprint arXiv:2408.11556},
  year={2024}
}

@article{fedus2022switch,
  title={Switch transformers: Scaling to trillion parameter models with simple and efficient sparsity},
  author={Fedus, William and Zoph, Barret and Shazeer, Noam},
  journal={Journal of Machine Learning Research},
  volume={23},
  number={120},
  pages={1--39},
  year={2022}
}

@inproceedings{jeon2019analysis,
  title={Analysis of $\{$Large-Scale$\}$$\{$Multi-Tenant$\}$$\{$GPU$\}$ clusters for $\{$DNN$\}$ training workloads},
  author={Jeon, Myeongjae and Venkataraman, Shivaram and Phanishayee, Amar and Qian, Junjie and Xiao, Wencong and Yang, Fan},
  booktitle={2019 USENIX Annual Technical Conference (USENIX ATC 19)},
  pages={947--960},
  year={2019}
}

@inproceedings{gupta2012study,
  title={A study of persistent threads style GPU programming for GPGPU workloads},
  author={Gupta, Kshitij and Stuart, Jeff A and Owens, John D},
  booktitle={2012 Innovative Parallel Computing (InPar)},
  pages={1--14},
  year={2012},
  organization={IEEE}
}

@article{dean2013tail,
  title={The tail at scale},
  author={Dean, Jeffrey and Barroso, Luiz Andr{\'e}},
  journal={Communications of the ACM},
  volume={56},
  number={2},
  pages={74--80},
  year={2013},
  publisher={ACM New York, NY, USA}
}

@inproceedings{chung2025swift,
  title = {Swift and Trustworthy Large-Scale GPU Simulation with Fine-Grained Error Modeling and Hierarchical Clustering},
  author = {Chung, Euijun and Na, Seonjin and Kang, Sung Ha and Kim, Hyesoon},
  booktitle = {Proceedings of the 58th IEEE/ACM International Symposium on Microarchitecture},
  pages = {1397--1411},
  year = {2025},
  artifact_available = {https://github.com/ejchung0406/STEM-AE},
  artifact_functional = {https://github.com/ejchung0406/STEM-AE}
}

@inproceedings{yu2022orca,
  title={Orca: A distributed serving system for $\{$Transformer-Based$\}$ generative models},
  author={Yu, Gyeong-In and Jeong, Joo Seong and Kim, Geon-Woo and Kim, Soojeong and Chun, Byung-Gon},
  booktitle={16th USENIX symposium on operating systems design and implementation (OSDI 22)},
  pages={521--538},
  year={2022}
}

@article{gim2024prompt,
  title={Prompt cache: Modular attention reuse for low-latency inference},
  author={Gim, In and Chen, Guojun and Lee, Seung-seob and Sarda, Nikhil and Khandelwal, Anurag and Zhong, Lin},
  journal={Proceedings of Machine Learning and Systems},
  volume={6},
  pages={325--338},
  year={2024}
}

@inproceedings{gao2024cost,
  title={$\{$Cost-Efficient$\}$ large language model serving for multi-turn conversations with $\{$CachedAttention$\}$},
  author={Gao, Bin and He, Zhuomin and Sharma, Puru and Kang, Qingxuan and Jevdjic, Djordje and Deng, Junbo and Yang, Xingkun and Yu, Zhou and Zuo, Pengfei},
  booktitle={2024 USENIX annual technical conference (USENIX ATC 24)},
  pages={111--126},
  year={2024}
}

@article{yao2022react,
  title={React: Synergizing reasoning and acting in language models},
  author={Yao, Shunyu and Zhao, Jeffrey and Yu, Dian and Du, Nan and Shafran, Izhak and Narasimhan, Karthik and Cao, Yuan},
  journal={arXiv preprint arXiv:2210.03629},
  year={2022}
}

@article{yuan2026kairos,
  title={KAIROS: Stateful, Context-Aware Power-Efficient Agentic Inference Serving},
  author={Yuan, Yichao and Chowdhury, Mosharaf and Talati, Nishil},
  journal={arXiv preprint arXiv:2604.16682},
  year={2026}
}

@article{nie2026queueing,
  title={A Queueing-Theoretic Framework for Stability Analysis of LLM Inference with KV Cache Memory Constraints},
  author={Nie, Chengyi and Si, Nian and Zhou, Zijie},
  journal={arXiv preprint arXiv:2605.04595},
  year={2026}
}

@article{liu2025contention,
  title = {Contention-Aware GPU Thread Block Scheduler for Efficient GPU-SSD},
  author = {Liu, Xueyang and Na, Seonjin and Chung, Euijun and Cao, Jiashen and Yang, Jing and Kim, Hyesoon},
  journal = {IEEE Computer Architecture Letters},
  year = {2025},
  publisher = {IEEE},
}

@article{zhu2026tracelab,
  title={TraceLab: Characterizing Coding Agent Workloads for LLM Serving},
  author={Zhu, Kan and Jacob, Mathew and Ma, Chenxi and Pan, Yi and Wang, Stephanie and Krishnamurthy, Arvind and Kasikci, Baris},
  journal={arXiv preprint arXiv:2606.30560},
  year={2026}
}

@article{wu2026dualpath,
  title={DualPath: breaking the storage bandwidth bottleneck in agentic LLM inference},
  author={Wu, Yongtong and Chen, Shaoyuan and Zhong, Yinmin and Huang, Rilin and Tan, Yixuan and Zhang, Wentao and Zhang, Liyue and Zhou, Shangyan and Liu, Yuxuan and Zhou, Shunfeng and others},
  journal={arXiv preprint arXiv:2602.21548},
  year={2026}
}

@article{ding2026observation,
  title={Observation, Not Prediction: Conversation-Level Disaggregated Scheduling for Agentic Serving},
  author={Ding, Jianru and Hosseini, Ryien and Gholami, Pouya Mahdi and Xiang, Mingyuan and Hoffmann, Henry},
  journal={arXiv preprint arXiv:2606.01839},
  year={2026}
}

@article{blakeman2025nvidia,
  title={Nvidia nemotron 3: Efficient and open intelligence},
  author={Blakeman, Aaron and Grattafiori, Aaron and Basant, Aarti and Gupta, Abhibha and Khattar, Abhinav and Renduchintala, Adi and Vavre, Aditya and Shukla, Akanksha and Bercovich, Akhiad and Ficek, Aleksander and others},
  journal={arXiv preprint arXiv:2512.20856},
  year={2025}
}
\end{document}